\documentclass[preprint,journal]{vgtc}       
\ifpdf
  \pdfoutput=1\relax                   
  \usepackage{graphicx}                
  \DeclareGraphicsExtensions{.pdf,.png,.jpg,.jpeg} 
\else
  \usepackage{graphicx}                
  \DeclareGraphicsExtensions{.eps}     
\fi%

\PassOptionsToPackage{dvipsnames}{xcolor}
\vgtcinsertpkg

\graphicspath{{figures/}{pictures/}{images/}{./}} 

\usepackage{url}
\usepackage[utf8]{inputenc}
\usepackage[T1]{fontenc}
\usepackage{microtype}                 
\PassOptionsToPackage{warn}{textcomp}  
\usepackage{textcomp}                  
\usepackage{mathptmx}                  
\usepackage{times}                     
\usepackage{cite}                      
\usepackage{booktabs}                  
\usepackage{cleveref}
\usepackage{t1enc}
\usepackage{xspace}
\usepackage{soul}
\usepackage{color}
\usepackage{relsize}
\usepackage[labelfont=sf]{subcaption}
\definecolor{orange}{rgb}{1,0.5,0}
\definecolor{green}{rgb}{0.5,0.8,0.3}
\newcommand{\Mspace}{M}
\newcommand{\Rspace}{R}

\hypersetup{colorlinks,
    citecolor=black,
    filecolor=black,
    linkcolor=black,
    urlcolor=black
}

\newcommand{\stwo}{1\xspace}
\newcommand{\sfour}{2\xspace}
\newcommand{\seight}{3\xspace}
\newcommand{\snine}{4\xspace}
\newcommand{\sten}{5\xspace}

\renewcommand{\paragraph}[1]{\noindent\textbf{#1}}

\newcommand{\showcomments}{}

\if10
\newcommand{\revision}[1]{\textcolor{blue}{#1}}
\newcommand{\revdel}[1]{\textcolor{red}{\st{#1}}}
\else
\newcommand{\revision}[1]{#1}
\newcommand{\revdel}[1]{}
\fi

\ifdefined\showcomments

\DeclareRobustCommand{\will}[1]{{\begingroup\sethlcolor{Orchid}\hl{(will:) #1}\endgroup}}
\DeclareRobustCommand{\steve}[1]
{{\begingroup\sethlcolor{CornflowerBlue}\hl{(steve:) #1}\endgroup}}
\DeclareRobustCommand{\torin}[1]
{{\begingroup\sethlcolor{green}\hl{(torin:) #1}\endgroup}}
\DeclareRobustCommand{\nate}[1]
{{\begingroup\sethlcolor{orange}\hl{(nate:) #1}\endgroup}}
\else

\DeclareRobustCommand{\will}[1]{}
\DeclareRobustCommand{\steve}[1]{}
\DeclareRobustCommand{\torin}[1]{}
\DeclareRobustCommand{\nate}[1]{}
\fi

\onlineid{1091}

\vgtccategory{Research}
\vgtcpapertype{application/design study}

\title{Improving the Usability of Virtual Reality Neuron Tracing with Topological Elements\vspace{-0.5em}}

\author{Torin McDonald, Will Usher, Nate Morrical, Attila Gyulassy, Steve Petruzza, \\
Frederick Federer, Alessandra Angelucci, and Valerio Pascucci}

\authorfooter{
\item Torin McDonald, Will Usher, Nate Morrical, Attila Gyulassy and Valerio Pascucci are with the SCI Institute, University of Utah.  torin@sci.utah.edu.
\item Steve Petruzza is with the SCI Institute, University of Utah and Utah State University.
\item Frederick Federer and Alessandra Angelucci are with the Moran Eye Institute, University of Utah.
}

\abstract{%
Researchers in the field of connectomics are working to reconstruct a map of neural
connections in the brain in order to understand at a fundamental level how the brain
processes information. Constructing this wiring diagram is done by tracing neurons
through high-resolution image stacks acquired with fluorescence microscopy imaging techniques.
While a large number of automatic tracing algorithms have been proposed, these frequently
rely on local features in the data and fail on noisy data or ambiguous cases,
requiring time-consuming manual correction. As a result, manual and semi-automatic
tracing methods remain the state-of-the-art for creating accurate neuron reconstructions.
We propose a new semi-automatic method that uses topological features to guide users in tracing neurons
and integrate this method within a virtual reality (VR) framework previously used
for manual tracing.  Our approach augments both visualization and interaction  with topological elements, allowing rapid understanding and tracing of complex morphologies. In our pilot study, neuroscientists demonstrated a strong preference for using our tool over prior approaches, reported less fatigue during tracing, and commended the ability to better understand possible paths and alternatives. Quantitative evaluation of the traces reveals that \revision{users' tracing speed increased, while retaining similar accuracy} compared to a fully manual approach.
}

\teaser{
  \centering
  \vspace{-0.5em}
  \begin{tabular}{@{}c@{\hskip 4pt}c@{\hskip 4pt}c@{\hskip 4pt}c@{}}
  
  \includegraphics[width=.24\linewidth]{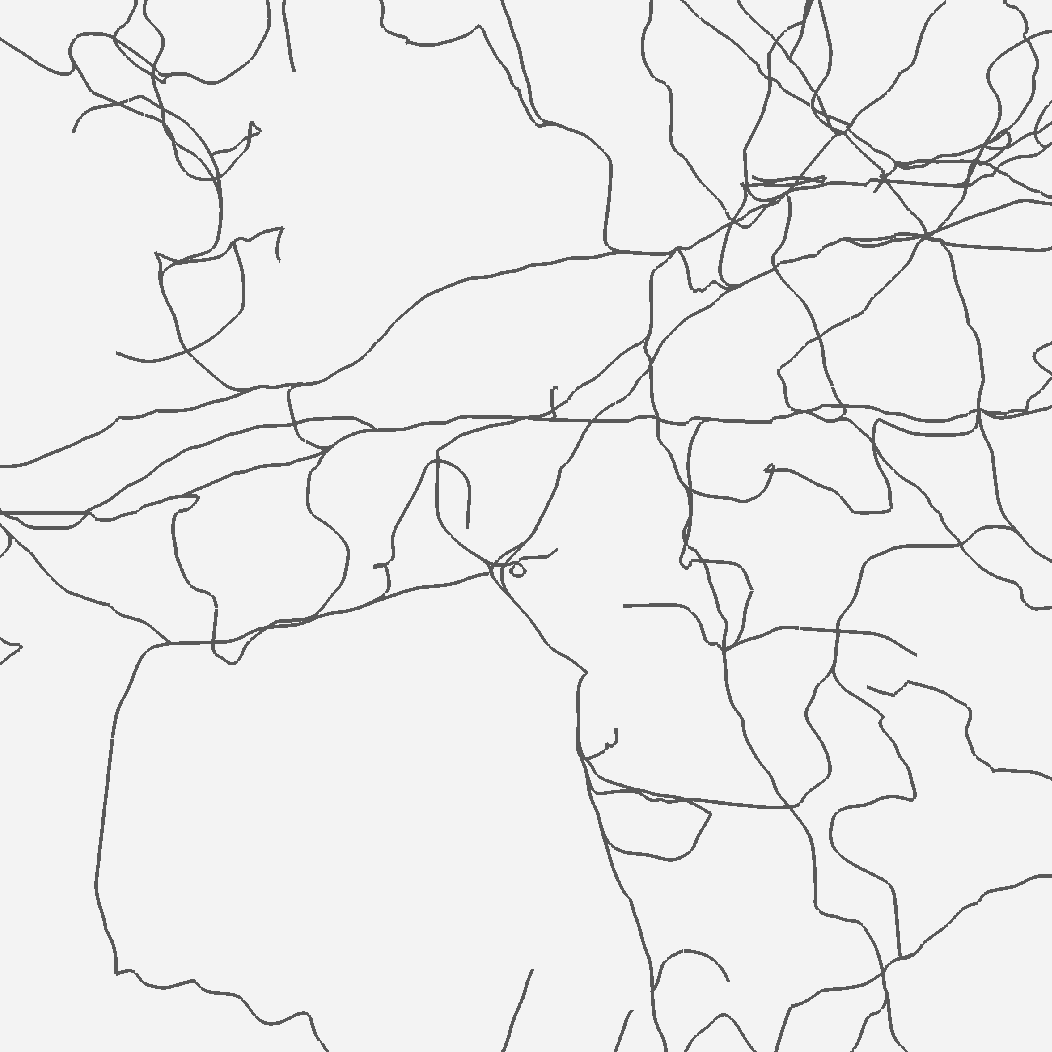} &
  \includegraphics[width=.24\linewidth]{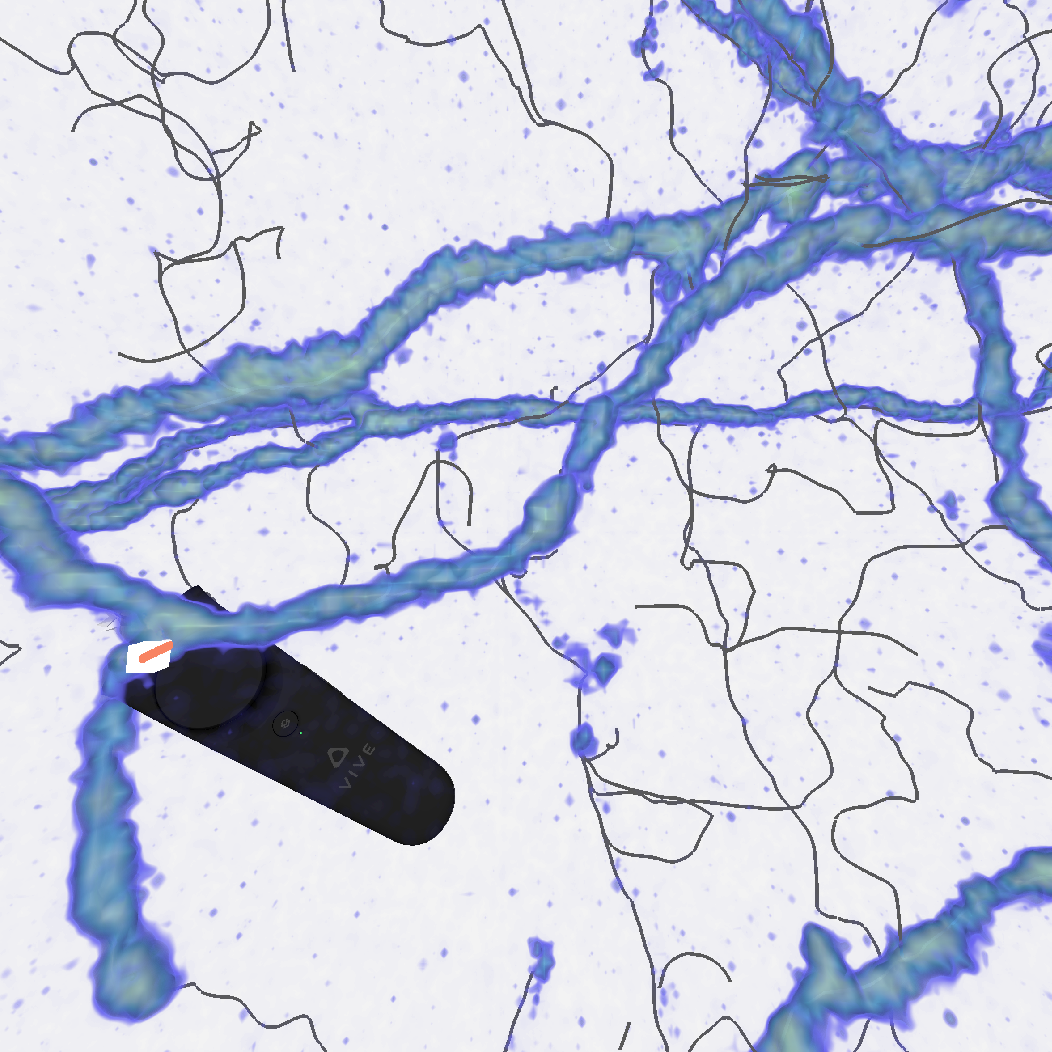} &
  \includegraphics[width=.24\linewidth]{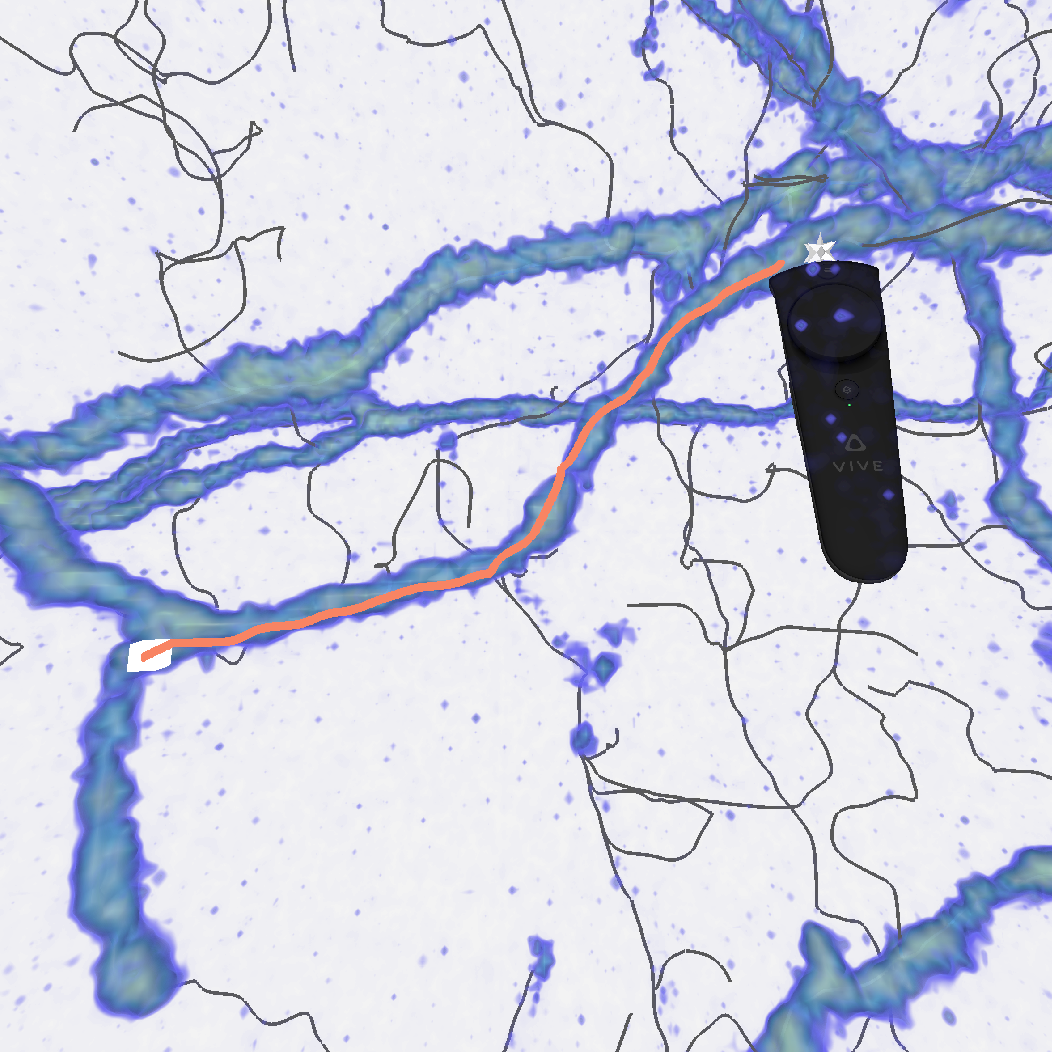} &
  \includegraphics[width=.24\linewidth]{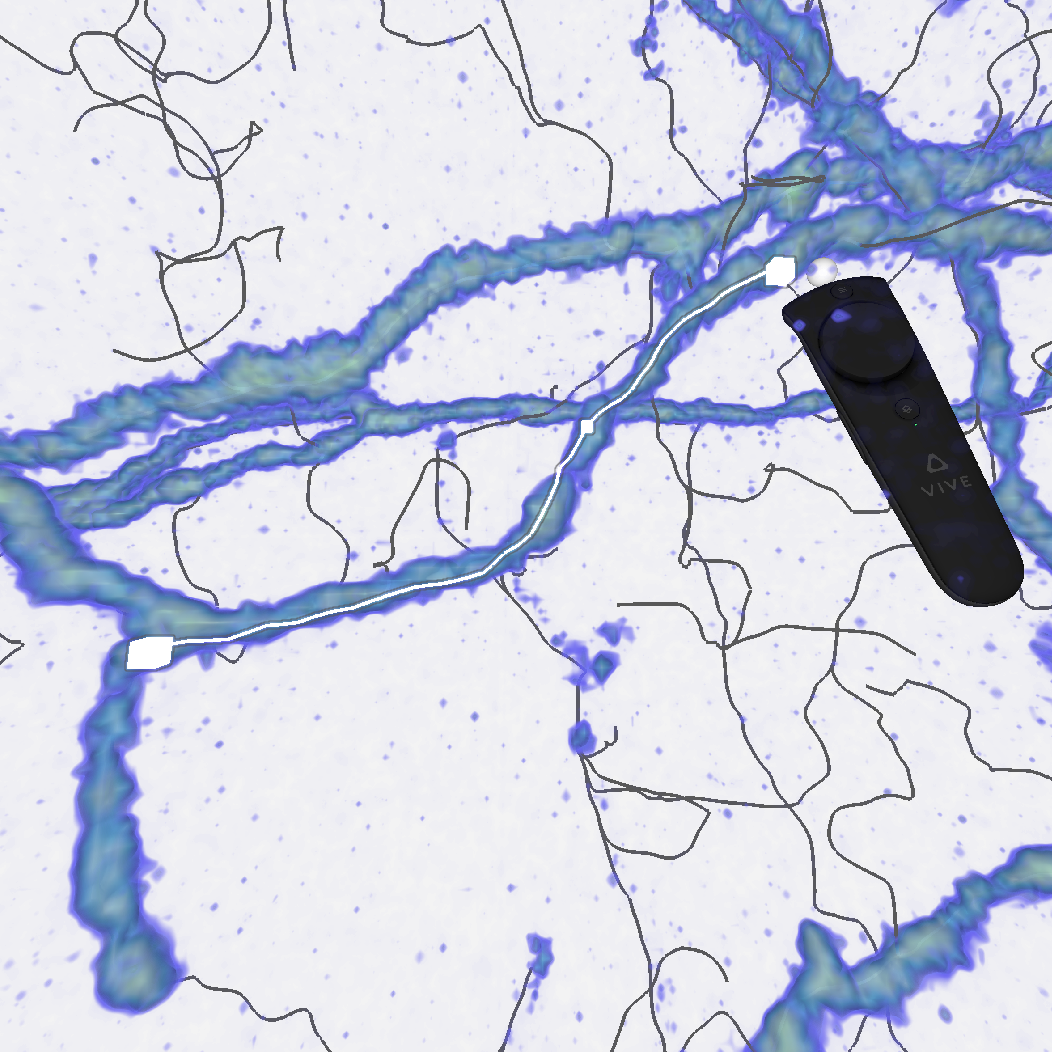} \\
  \end{tabular}
  \vspace{-1em}
  \caption{\label{fig:teaser}%
	Left to right: A connected graph of ridge-like structures is extracted from the Morse-Smale complex \revision{(MSC)},
	containing a superset of the possible neuron segments in the data.
	Our MSC-guided semi-automatic tracing tool enables users to rapidly trace paths and view a live preview as they do so (orange line).
	When satisfied with the trace, they can add it to the reconstruction (white line).
	}
}

\begin{document}


\maketitle
\section{Introduction}
A central goal within the field of neuroscience is to understand how
the dense, interconnected neural circuits in the brain communicate
and process information, and how this processing relates to behavior.
The field of connectomics was founded to understand
the fundamental wiring map of the brain in order to comprehend these neural circuits
at a mechanistic level.
Through analyzing neuron structure and connectivity, neuroanatomists
can gain a deeper understanding of fundamental brain functions
and new insights about brain diseases and treatments. 


However, obtaining a comprehensive wiring diagram for even relatively small and simple
mammalian brains, such as that of a mouse, is a massive
undertaking~\cite{bock_network_2011,briggman_wiring_2011,oh_mesoscale_2014,silvestri_confocal_2012}.
Projects focusing on species with larger brains more similar to humans,
such as non-human primates (NHP), are even more challenging.
Although recent advancements in
high-resolution tissue labeling~\cite{luo_genetic_2008},
optical tissue clearing~\cite{chung_structured_2013,yang_pact_2014,murray_switch_2015} and
imaging~\cite{silvestri_confocal_2012,oh_mesoscale_2014} have made it possible to image NHP
brains at large scales and high resolutions, the technology for extracting the imaged
neuron morphologies has struggled to keep up.

Current efforts to improve the speed of neuron morphology extraction have largely focused on fully automatic techniques. Automatic techniques take
a stack of images and attempt to extract the imaged neuron structures,
without user input.
The DIADEM (DIgital reconstructions of Axonal and DEndritic Morphology)
Challenge~\cite{gillette_diadem_2011} was proposed in 2009 to motivate improvement
of these techniques. The ultimate goal of this community effort was to increase
the speed that neurons could be traced by $20\times$. However,
at the end of the challenge no algorithm had achieved this goal due to
the laborious post-processing required to correct errors~\cite{liu_diadem_2011}.
Peng et al.~\cite{peng_proof-editing_2011} reported that this
post-processing step can take longer than a manual tracing.
Although additional efforts to improve automatic reconstruction are ongoing~\cite{peng_diadem_2015},
in practice the bulk of neuron tracing is done manually~\cite{meijering_neuron_2010} or with
a semi-automatic method.



Manually tracing neurons is a difficult and time-consuming process.
Tracing is typically done on a desktop, using standard software (e.g., NeuroLucida~\cite{neurolucida},
Vaa3D~\cite{peng_v3d_2010}). The data is displayed as a 2D set of images or 3D volume,
and the user clicks along the neuron to draw a \revision{path}. The lack of ability to directly
make selections in 3D or navigate the data in 3D introduces additional usability
challenges on top of the already difficult task of tracing.
To address this issue, Usher et al.~\cite{usher2018virtual} proposed a Virtual Reality (VR)
based tool for manual neuron tracing and found that neuroscientists using
the tool performed similar quality traces in less time.



Semi-automatic neuron tracing methods have been proposed to provide a compelling alternative
to both manual and fully automatic neuron tracing~\cite{neurolucida_360,peng_v3d_2010,peng2010automatic,myatt_neuromantic_2012}.
When using a semi-automatic method,
the user provides coarse guidance to the algorithm, e.g., through a set of start and end points
or clicks. The algorithm then extracts the neuron structure between these guide points.
Semi-automatic methods can
significantly reduce the amount of time taken to trace a neuron by integrating the
neuroscientist's guidance into the algorithm, reducing the amount of post-processing manual cleanup
required.

In this work we propose a new semi-automatic neuron tracing framework that builds
\revision{on} topological analysis methods~\cite{MSCEER}, developed through direct collaboration
with expert neuroanatomists.
Our approach uses the Morse-Smale complex \revision{(MSC)} to 
precompute a superset of potential paths that follow neurons. Having access to this superset 
of traces allows neuroanatomists to quickly trace along the neuron of interest
by selecting subsets of these paths.
We implement our semi-automatic method
within a virtual reality neuron tracing system to provide an intuitive environment
to work with the 3D data.
In a pilot study with neuroanatomists, we find that our approach provides significant
benefits, retaining
trace accuracy while improving speed and reducing fatigue. Moreover, the additional guides provided
by our tool assist the user in interpreting the data.
Our contributions are:
\begin{itemize}
    \setlength\itemsep{0em}
	\item A novel topologically guided framework for real-time semi-automatic neuron tracing
	\item An intuitive interaction design for using this framework in VR
	\item A comparison of our approach against widely used semi-automatic methods 
	as well as previous manual tracing methods through a user study with domain experts.
\end{itemize}

\section{Background and Related Work} 
	
	
	
The neuron morphology reconstruction workflow has a number of components, with one of the most time consuming being the physical tracing of neurons.
To provide context for neuron tracing, we describe the typical reconstruction workflow in practice (\Cref{sec:workflow}).
We then review current automatic and semi-automatic neuron tracing methods and
their limitations (\Cref{sec:auto_semiauto}) and the state of the art in immersive environments (\Cref{sec:immersive}).
Finally, we review the Morse-Smale complex and 
examples of its application to analysis tasks in other scientific domains (\Cref{sec:MSC}).

\subsection{Neuron Tracing Workflow}
\label{sec:workflow}



Modern methods for acquiring neuron microscopy data use viral vectors
carrying genes for fluorescent proteins~\cite{luo_genetic_2008}. When injected
into the tissue these vectors induce fluorescence within the structures to be
imaged, labelling them at high resolution.
The brain tissue is then rendered optically transparent using a clearing
technique such as CLARITY~\cite{chung_structured_2013}, PACT~\cite{yang_pact_2014}, or SWITCH~\cite{murray_switch_2015},
and imaged in blocks with a confocal or two-photon microscope.
These methods allow for imaging large blocks of tissue or entire brains, and
can produce terabytes of high-resolution image stacks.

To reconstruct the labeled neurons from these image stacks, neuroanatomists
use commercial tools like NeuroLucida~\cite{neurolucida} or 
open-source tools like Vaa3D~\cite{peng_v3d_2010}. These tools display the collected image stacks
as either a set of 2D slices or as a 3D volume,
where the user can trace manually by drawing lines along the structures of interest,
or guide a semi-automatic algorithm along the structures to extract them.
Once the desired neurons have been reconstructed, they can be used in
brain function simulations or overlaid on functional maps of the
brain, to understand the connectivity between brain regions.
Although fully automatic algorithms are supported by standard tools,
they are less widely used in practice due to issues with image quality
and ambiguity. \revision{Our coauthor's lab employs} several trained undergraduates
responsible for the bulk of the neuron tracing work, with additional
tracing done by graduate students and research scientists.



\subsection{Automatic and Semi-Automatic Neuron Tracing} 
\label{sec:auto_semiauto}
Today, neuron tracing remains a crucial bottleneck in the field
of connectomics~\cite{meijering_neuron_2010}. A large body of work has been
devoted to developing new methods to accelerate this process,
either through fully automatic algorithms or semi-automatic user-guided algorithms.

A significant ongoing effort in the community has been to develop and
evaluate fully automatic algorithms for neuron reconstruction. Two community
efforts, the DIADEM Challenge~\cite{gillette_diadem_2011} and the ongoing
BigNeuron Project~\cite{peng_diadem_2015,peng_bigneuron_2015},
seek to provide a test bed for evaluating new 
reconstruction algorithms. Results from the DIADEM 
challenge suggest that the current state-of-the-art 
automatic tracing algorithms are not suitable for widespread use in practice, because
significant manual post-processing is required~\cite{peng_proof-editing_2011}.
\revision{Recent work has begun applying
Machine Learning techniques to the neuron segmentation problem}~\cite{soltanian-zadeh_fast_2019,apthorpe_automatic_2016};
\revision{however, as with other ML-based approaches, a large amount of training data must
be provided as input to the algorithm, that must be produced using existing techniques.
In contrast to work segmenting space-filling neurons imaged using electron microscopy}~\cite{jeong2009scalable},
\revision{our approach works on sparsely labeled neurons as linear structures.}
For a full review of recent
advances we refer to the \revision{recent survey by Magliaro et al.}~\cite{magliaro_gotta_2019}
and that by Acciai et al.~\cite{acciai_automated_2016}. 

Due to the challenges in using fully automatic methods in practice,
semi-automatic algorithms have found a growing interest in the community. When using
a semi-automatic reconstruction algorithm, the user guides the algorithm
along the neuron by tracing roughly along the neuron or
clicking to mark start, branch, and end points to connect. By integrating
more guidance from the neuroscientist into the algorithm, the amount
of additional post-processing cleanup required can be reduced,
while still decreasing the time spent tracing compared to a fully manual trace.
For example, Vaa3D's semi-automatic approach uses
a pixel based shortest path algorithm~\cite{peng2010automatic} to connect the start point and one
or more markers placed by the user. NeuroLucida 360's~\cite{neurolucida_360} semi-automatic tracing works similarly,
where the user traces along the feature to guide the algorithm to important features.
Neuromantic~\cite{myatt_neuromantic_2012}
uses a 3D extension of Meijering et al.'s 2D steerable
Gaussian filter algorithm~\cite{meijering2004design} for semi-automatic reconstruction.

However, these methods all work in the context of traditional desktop software,
taking 2D inputs from a mouse and providing 2D imagery through a monitor.  
For example, Vaa3D's \emph{Virtual Finger}~\cite{peng_virtual_2014} casts
rays through the volume to find the potentially selected objects as the user
draws a line with the mouse.
Thus, users may need to perform multiple interactions and camera rotations
to find and select the desired feature,
to work around occluders or ambiguous hits in the ray casting process.
Furthermore, such methods typically operate on the underlying image data and
are computationally intensive, thereby impacting interactivity.

\subsection{Immersive Environments}
\label{sec:immersive}

There has been a growing interest in using virtual reality or immersive environments
for neuron tracing and visualization in general to overcome the limitations of traditional
2D desktop interaction and visualization modalities. Existing tools such
as Vaa3D have announced early VR system support, and other new VR-specific tools
have been released~\cite{usher2018virtual,syglass}.
In contrast to desktop software, VR and immersive systems allow users to visualize and interact
with their data directly in 3D, providing a more intuitive
interface and allowing for better understanding of 3D
structures~\cite{prabhat_comparative_2008,laha_effects_2014,laha_effects_2012}.

Usher et al.~\cite{usher2018virtual} proposed a virtual reality system
for \revision{room scale or seated} manual neuron tracing. In their evaluation, they found that domain experts
could perform similar quality traces to standard desktop software in less time,
achieving a roughly $2\times$ speedup. Moreover, they found that experts reported
the VR tool to be more intuitive and less fatiguing, with the immersive
visualization aiding their understanding of the data. However, their tool supports only
manual tracing and thus, while faster than working on a desktop, would still require
a significant amount of time and effort to trace large data sets.

Immersive systems have also been proposed for visualization of electron and
wide-field microscopy data. 
\revision{Agus et al.}~\cite{agus_glam_2018} \revision{proposed a model for simulating lactate absorption to
compute lactate absorption maps on 3D segmented electron microscopy data sets.
The absorption maps can then be visualized in a VR or CAVE environment, by rendering
the segmented neuron meshes colored by absorption rate.}
\revision{Boges et al.}~\cite{boges_immersive_2019,boges_virtual_2020} \revision{proposed a
virtual reality tool for creating, editing, and visualizing skeletonizations
of brain cells in electron microscopy data, along with their segmented surface mesh.
The neuron skeleton is created by the user with the assistance of a guidance system that
moves their points placed inside the mesh to its center.
In the use case we target, a segmentation of the data is not available.
Instead, we leverage the Morse-Smale complex to guide users when tracing through a volumetric representation.}
Boorboor et al.~\cite{boorboor_visualization_2019} proposed
a data processing and feature extraction pipeline, the output of which
could be visualized in an immersive display wall visualization system implemented with
Unity. Sicat et al.~\cite{sicat_dxr_2019} presented DXR, a Unity based toolkit for easily developing
immersive visualization applications. Fulmer et al.~\cite{fulmer_imweb_2019} presented
a web-based immersive neuron visualization system using Unity to explore
online databases of neuron data in a Hololens.



\subsection{Topological Analysis}
\label{sec:MSC}
Topological methods have been shown to be highly effective in extracting
application-specific features of interest (e.g.,~\cite{bock_topoangler_2018,
rieck_clique_2018,TopoMS,Laney06vis, Gyulassy07btvcg,Sousbie2011,
GKLW16,Bremer09tvcg,weber_topology-controlled_2007}).
In many cases,
the features of interest can be defined directly in terms of topological structures
such as the Contour Tree or the Morse-Smale complex (MSC)\@.
Recent advances in the computation of the MSC~\cite{Robins11, Shivashankar2012} have
made it readily available to the broader scientific community through open source libraries
such as TTK~\cite{ttk} and MSCEER~\cite{MSCEER}.
In this work, we build our framework for neuron tracing on top of the MSC\@.

Given a compact $d$-manifold $\Mspace$, a scalar function $f: \Mspace \rightarrow \Rspace$ is a \textit{Morse} function if its \textit{critical points} are non-degenerate and have distinct values. A
critical point occurs where the gradient vanishes, $\nabla f = 0$, and is non-degenerate if its Hessian is non-singular. 

\revision{For three-dimensional domains, a critical point is either a \textit{minimum}, \textit{1-saddle}, \textit{2-saddle}, or \textit{maximum}.}
%
%
An integral line in $f$ is a path in $\Mspace$ whose tangent vector is parallel to
the gradient of $f$ at each point along the path. The lower limit of the integral
line is called the \textit{origin}, and the upper limit the \textit{destination}.
These lower and upper limits occur at critical points of $f$. \emph{Ascending} and
\emph{descending} manifolds are obtained as clusters of integral lines having common origin
and destination, respectively. The descending manifolds of $f$ partition $\Mspace$ into a
cell complex called the \textit{Morse} complex. Symmetrically, the ascending manifolds
also partition $\Mspace$ into a cell complex.
A Morse function $f$ is a \textit{Morse-Smale function} if the ascending and descending manifolds
of its critical points intersect only transversally. The 0- and 1-dimensional cells of the
intersection of ascending and descending manifolds form the \textit{1-skeleton} of the Morse-Smale complex.
Practically, the MSC \textit{1-skeleton} is formed by \textit{nodes} and \textit{arcs}.
\textit{Nodes} are the critical points of the MSC, and \textit{arcs} are
the integral lines connecting critical points which differ in index by one.
For a complete visual overview of the components of the MSC we refer
to Gyulassy et al.~\cite{gyulassy2012direct}.

\begin{figure}[t]
	\centering
	\begin{subfigure}{0.32\columnwidth}
		\centering
		\includegraphics[width=\textwidth]{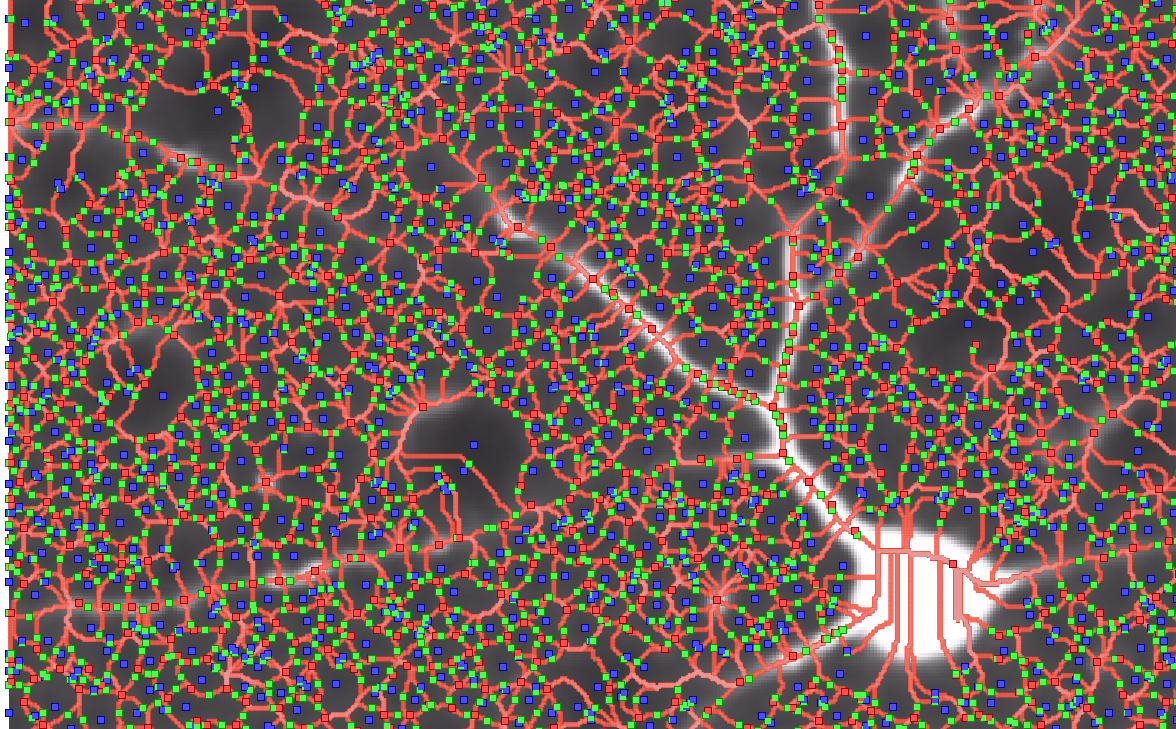}
	\end{subfigure}
	\begin{subfigure}{0.32\columnwidth}
		\centering
		\includegraphics[width=\textwidth]{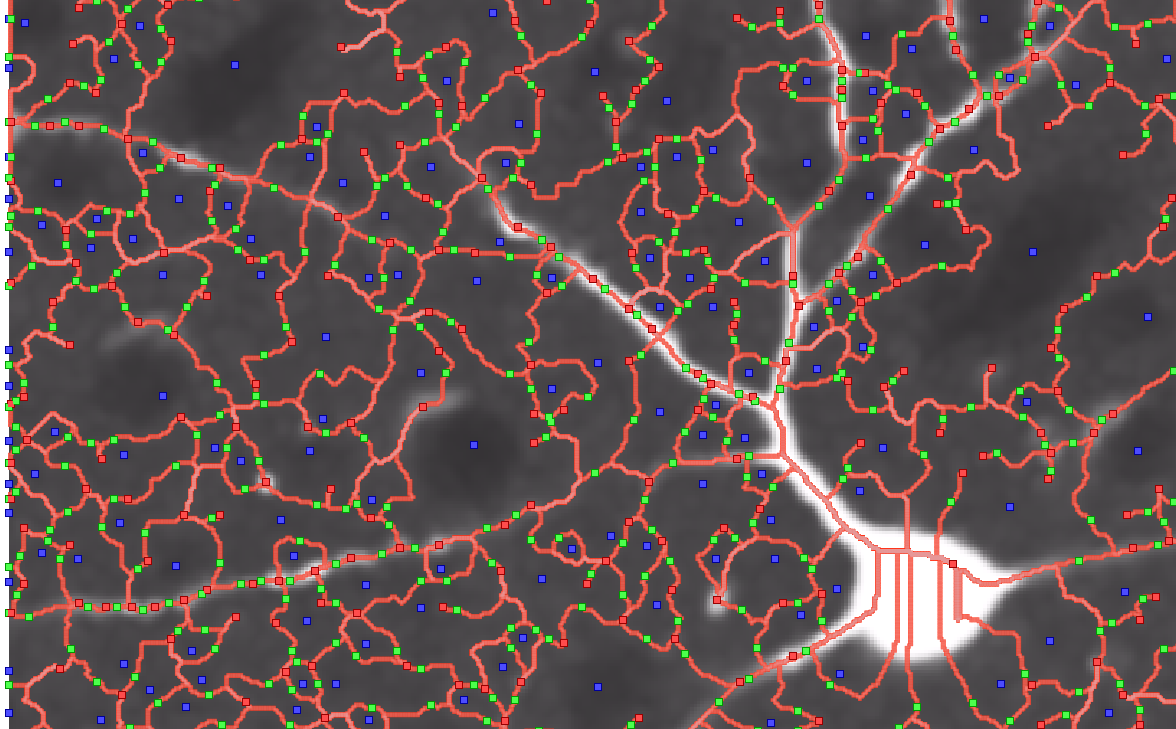}
	\end{subfigure}
	\begin{subfigure}{0.32\columnwidth}
		\centering
		\includegraphics[width=\textwidth]{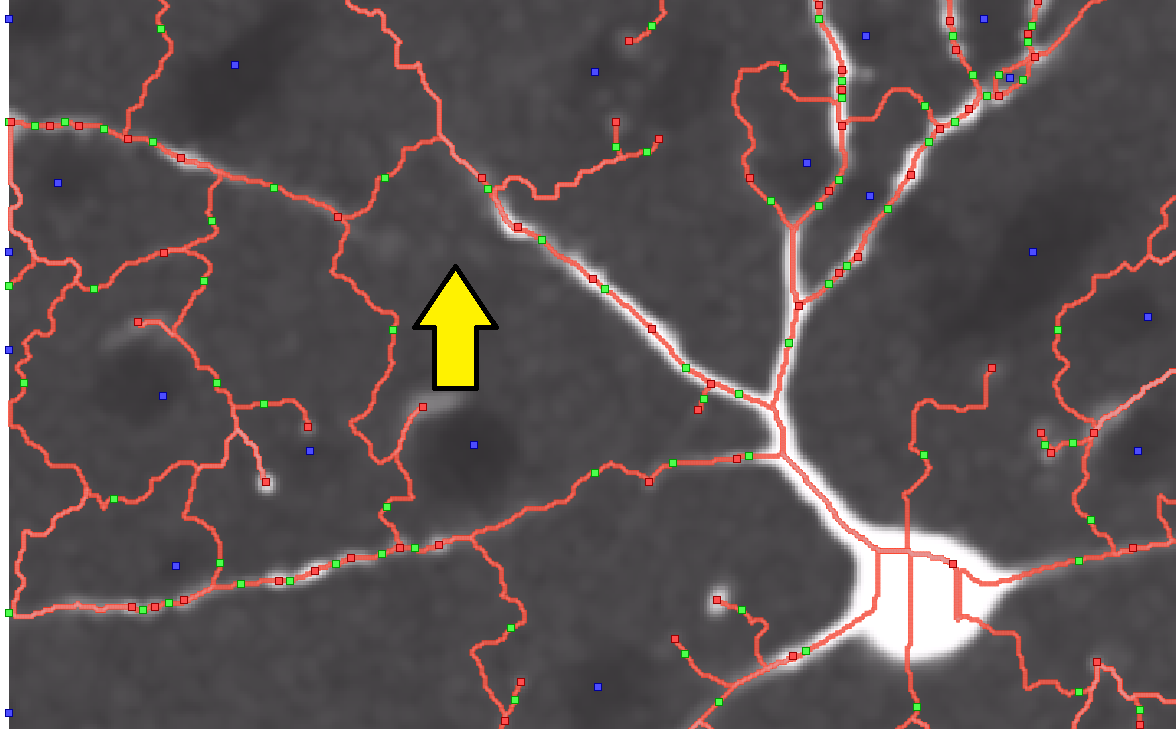}
	\end{subfigure}
	\vspace{-0.5em}
	\caption{\label{fig:mscgraph}%
	From left to right: \revision{the saddle-maximum arcs of the MSC are simplified with increasing
	persistence thresholds. Red lines represent saddle-maximum arcs, and red and green points correspond to maxima and saddles, respectively. A subset of the arcs cover neurons (left, middle), however, over-simplification discards faint features (right).}}
	\vspace{-1.5em}
\end{figure}

One of the greatest advantages of using a topological approach is the multiscale analysis 
made possible by persistent homology~\cite{Edelsbrunner2000}, which allows for the simplification
of noisy topological features.
Persistence simplification repeatedly removes critical point pairs that form the birth and
death of a topological feature, based on the lifespan of the feature in a sweep of the range of the function.
Low-intensity noise often appears as low-persistence features that can be removed while maintaining an overall connected structure. 
The MSC 1-skeleton has well understood rules that govern its persistence simplification
through repeatedly removing pairs of critical points connected by arcs,
and reconnecting their neighborhoods~\cite{Gyulassy2006}.
Our use of the MSC is motivated by the observation that the ridge-like structures
formed by the 1-skeleton of the simplified MSC, composed of the arcs between 2-saddles and maxima, correspond to the center
lines of the vast majority of neurons in the data, as shown in~\Cref{fig:mscgraph}. 

The components of the MSC have been used in practice to extract features
of interest in a range of application domains.
For example, these components define features in the electron density field in the
quantum theory of atoms in molecules: maxima occur at atom locations; 2-saddle-maximum arcs are
covalent bonds; and descending 3-manifolds are atomic basins~\cite{TopoMS}.
In other domains, specially selected subsets of the MSC can be used to extract features:
descending 2-manifolds represent bubbles in mixing fluids~\cite{Laney06vis};
2-saddle-maximum arcs can be used to extract the core of a porous solid~\cite{Gyulassy07btvcg} as well as the filamentary structure of galaxies~\cite{Sousbie2011} or structural materials~\cite{petruzza2019high};
descending 2- and 1-manifolds identify lithium diffusion pathways~\cite{GKLW16};
and ascending 2-manifolds define burning regions in combustion simulations~\cite{Bremer09tvcg}.
In each application, the features of interest were computed by identifying the appropriate topological
abstraction, and then selecting a subset of the topological features that correspond to the
quantities under study. 


While our work is inspired by these approaches, the images generated from fluorescence microscopy pose
a massive challenge for automated analysis. In addition to high-intensity noise, images of neurons have uneven staining, shadows, alignment artifacts \revision{from stitching image tiles}, and other unexplained gaps in the signal which require manual intervention.
This poses a challenge to topological methods that report what is in the
scalar function itself, faithfully representing artifacts and noise along with the desired signal.

\section{Topology Guided Neuron Tracing}
A semi-automatic method for neuron tracing in a VR environment faces two main constraints.
First, maintaining interactivity with the high framerate requirements of VR places a hard time constraint on point queries, neighborhood queries, and path computation.
As image volumes can reach tens to hundreds of gigabytes in size, on-the-fly computation on the raw voxel data is unable to meet these constraints.
Second, the image quality varies dramatically even within a single volume.
Traditional skeletonization and ridge-extraction techniques that take a background or foreground threshold will either over- or under-connect features of interest within the volume.
Our topology-based ridges meet both requirements: (i) a fast and scalable precomputation produces a sparse data structure that is fast to query \revision{when tracing} and (ii) the MSC-graph produced adapts to local image quality without relying on a \revision{global value threshold.}


\subsection{Computing MSC-Graphs} 

In the images produced through the imaging process described
in \Cref{sec:workflow}, high-intensity values correspond to the labeled \revision{cell bodies (soma),
and the structures connecting them to each other (dendrites and axons), which form each neuron.}
When tracing these structures, the user aims to
produce a path that follows the center line of these ridge-like structures.
Our approach in this work is to generate \textit{every possible ridge-like path} first,
turning the neuron reconstruction task into an efficient and interactive subset selection on a sparse data structure.
This is in sharp contrast to existing semi-automatic and automatic methods, that attempt
to mimic the manual extraction process by computing the single most-likely
path for the user~\cite{peng2011automatic,myatt_neuromantic_2012,yang_fmst_2018}.

\begin{figure}[t]
 	\centering
 	\includegraphics[width=\linewidth]{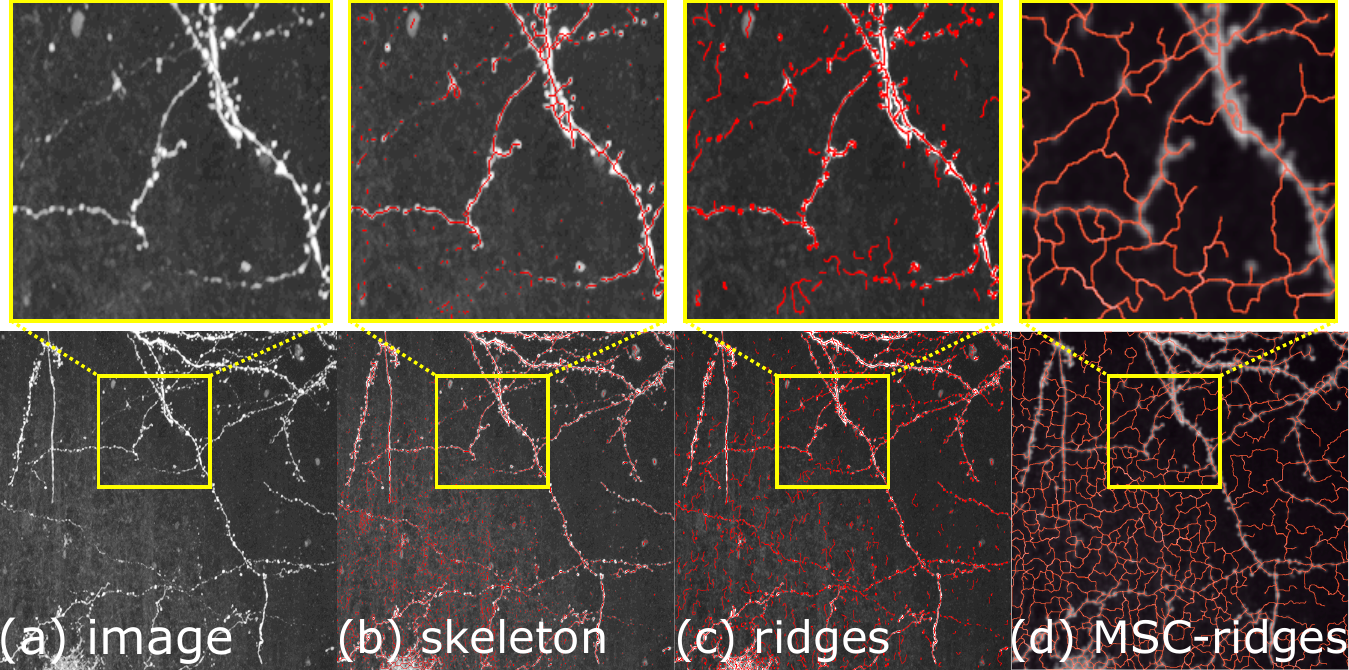}
	\vspace{-1.5em}
	\caption{\label{fig:msccompare}%
	Comparison of ridge structure computation. Both (b) skeletonization through iterative thinning~\cite{Lee94} and (c) second-derivative ridge detection~\cite{Steger98} have gaps in the structure, whereas (d) the MSC-graph~\cite{gyulassy2012direct} produces connected paths. Parameters were adjusted in (b) and (c) to maximally connect the network; however, neither approach was able to recover from gaps in the signal or reproduce faint axons.}
	\vspace{-1em}
\end{figure}

Our first task is to extract the set of all possible ridge lines from
the scalar field.
Historically, ridge lines have been defined with techniques relating to the alignment of
the principle directions of curvature and the gradient, Eberly et al.~\cite{Eberly94} provide an excellent overview.
However, locally defined ridge lines have major limitations for the task of acting as an acceleration structure
for neuron reconstruction.
Height ridges do not necessarily form an interconnected network, with segments ending where the local image no longer looks like a ridge. Pruning ridge lines by intensity further disconnects the network, exacerbating the problem. 
Instead, we use a topological approach based on the MSC that identifies ridge-like structures that are close enough to true ridge lines, but also form an interconnected network ideal for path computations. Figure~\ref{fig:msccompare} compares ridge-lines computed with iterative thinning, second-derivative ridges, and the MSC. 

 Although integral lines do not merge for continuous functions, the computational methods utilize discrete Morse theory~\cite{Forman:2002, Gyulassy2019}, where 2-saddle-maximum paths may merge. Our data structure inserts a new node at each merger to maintain the property that any vertex along a path can be mapped to an arc of the data structure. In the remainder of this text, we use \textit{MSC-graph} to refer to this modified 1-skeleton.
\Cref{fig:mscgraph} illustrates the full MSC and its successive coarsening through 
persistence simplification.
The steps for computing our data structure are image preprocessing, MSC computation and persistence simplification, dividing arcs to form the MSC-graph, and geometric smoothing of the MSC-graph.

\begin{figure}[t]
	\centering
	\begin{subfigure}{0.49\columnwidth}
		\centering
		\includegraphics[width=.49\textwidth]{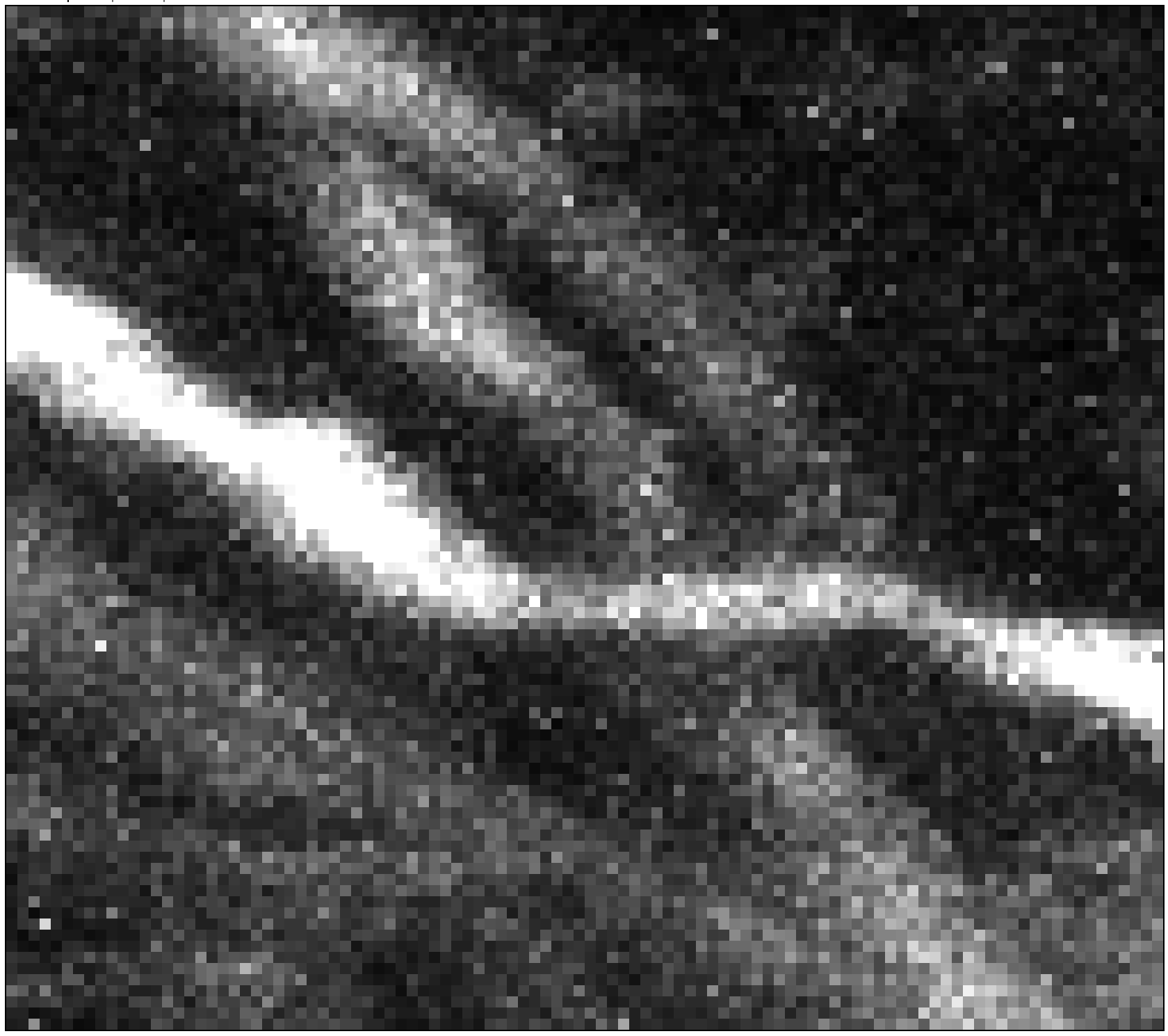}\hfill
		\includegraphics[width=.49\textwidth]{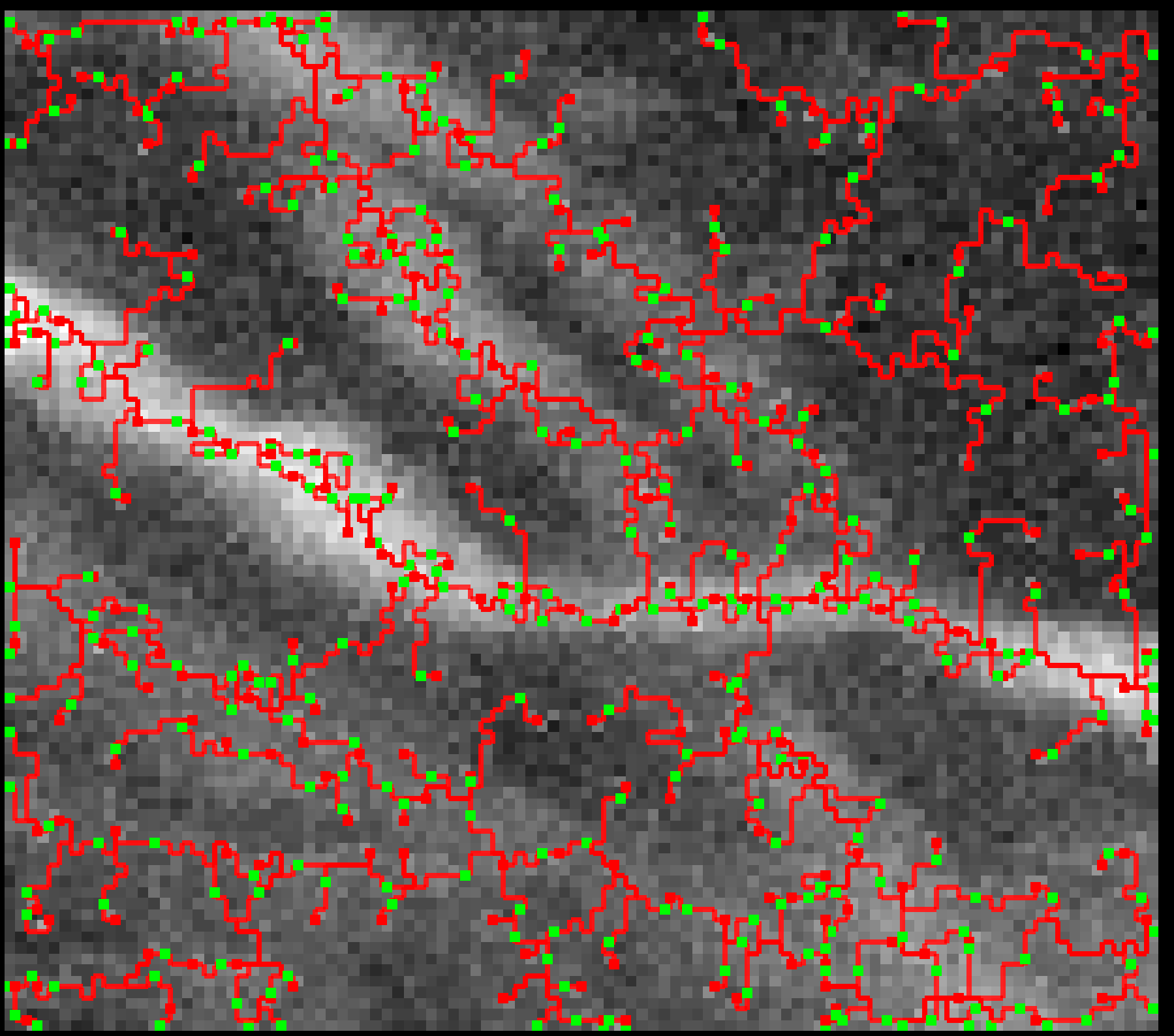}
		 \caption{\label{fig:noisy_and_rg}%
		 The original image data.}
	 \end{subfigure}
	\begin{subfigure}{0.49\columnwidth}
		\centering
		\includegraphics[width=.49\textwidth]{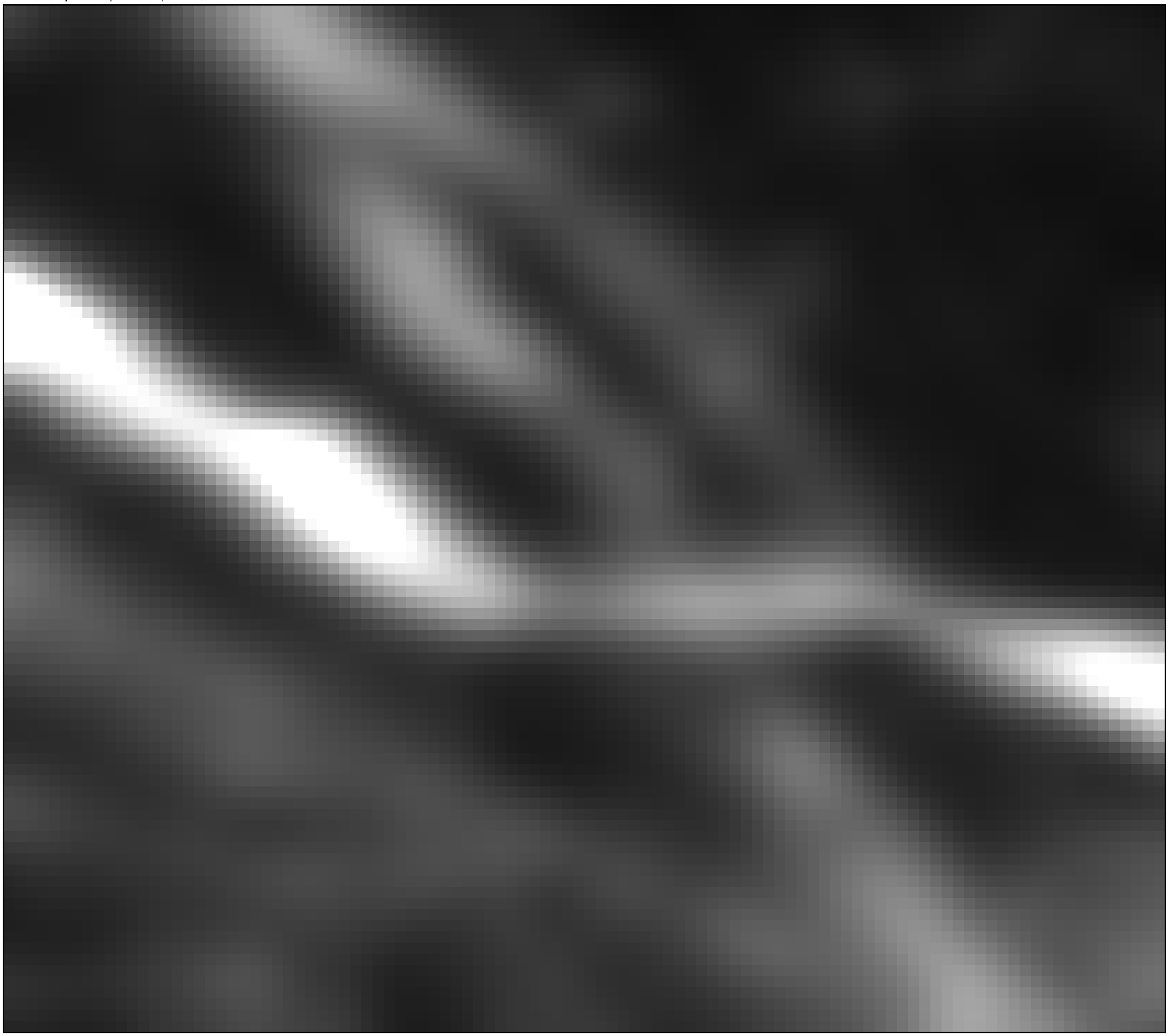}\hfill
		\includegraphics[width=.49\textwidth]{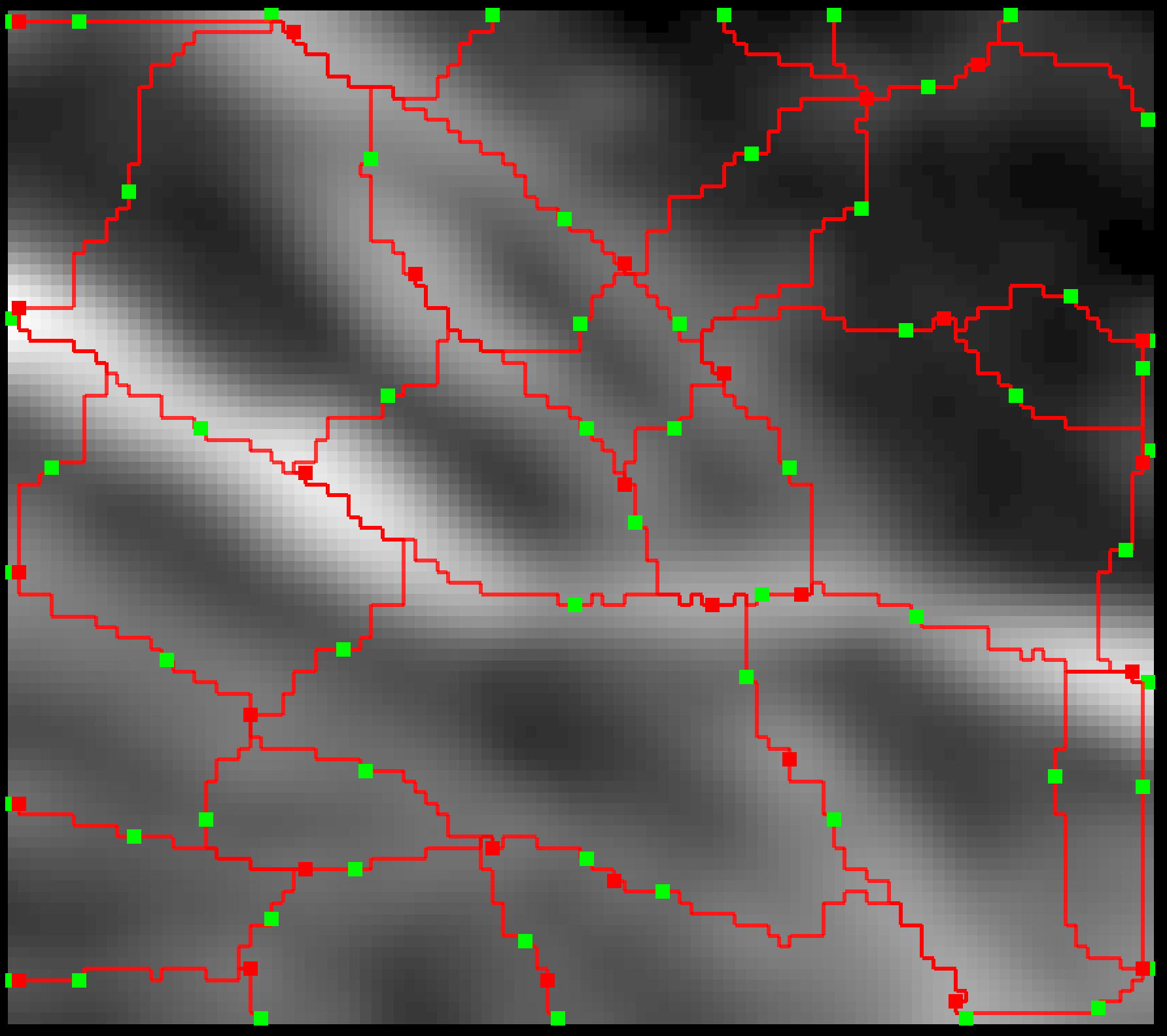}
		\caption{\label{fig:smooth_and_rg}%
		After filtering and blurring.}
	\end{subfigure}
	\vspace{-1em}
	\caption{\label{fig:improc}%
	The image preprocessing step reduces the effect of noise,
	creating a sparser initial MSC-graph with a better geometric embedding.}
	\vspace{-1.5em}
\end{figure}

\paragraph{Image Preprocessing.}
\revision{Noise in microscopy images leads to poor geometric embedding of the MSC-graph,  since it is sensitive to discontinuities in the image gradient (\Cref{fig:noisy_and_rg}). The microscopy images used in this study show characteristic \textit{speckle noise}, high-frequency, high-amplitude intensity spikes. A common practice is to first remove noise before further analysis~\cite{acciai_automated_2016}. While many sophisticated approaches exist for noise removal, a simple combination of median and Gaussian blur filters (with radius 2) led to MSC-graphs with both fewer arcs and better embedding (\Cref{fig:smooth_and_rg}). This filter combination produced excellent results for all images in the study coming from different staining techniques, tissue samples, magnifications, and microscopes. However, different acquisitions may require alternative denoising approaches.}   



\begin{figure}[t]
	\centering
	\begin{subfigure}{0.48\columnwidth}
		\centering
		\includegraphics[width=\textwidth]{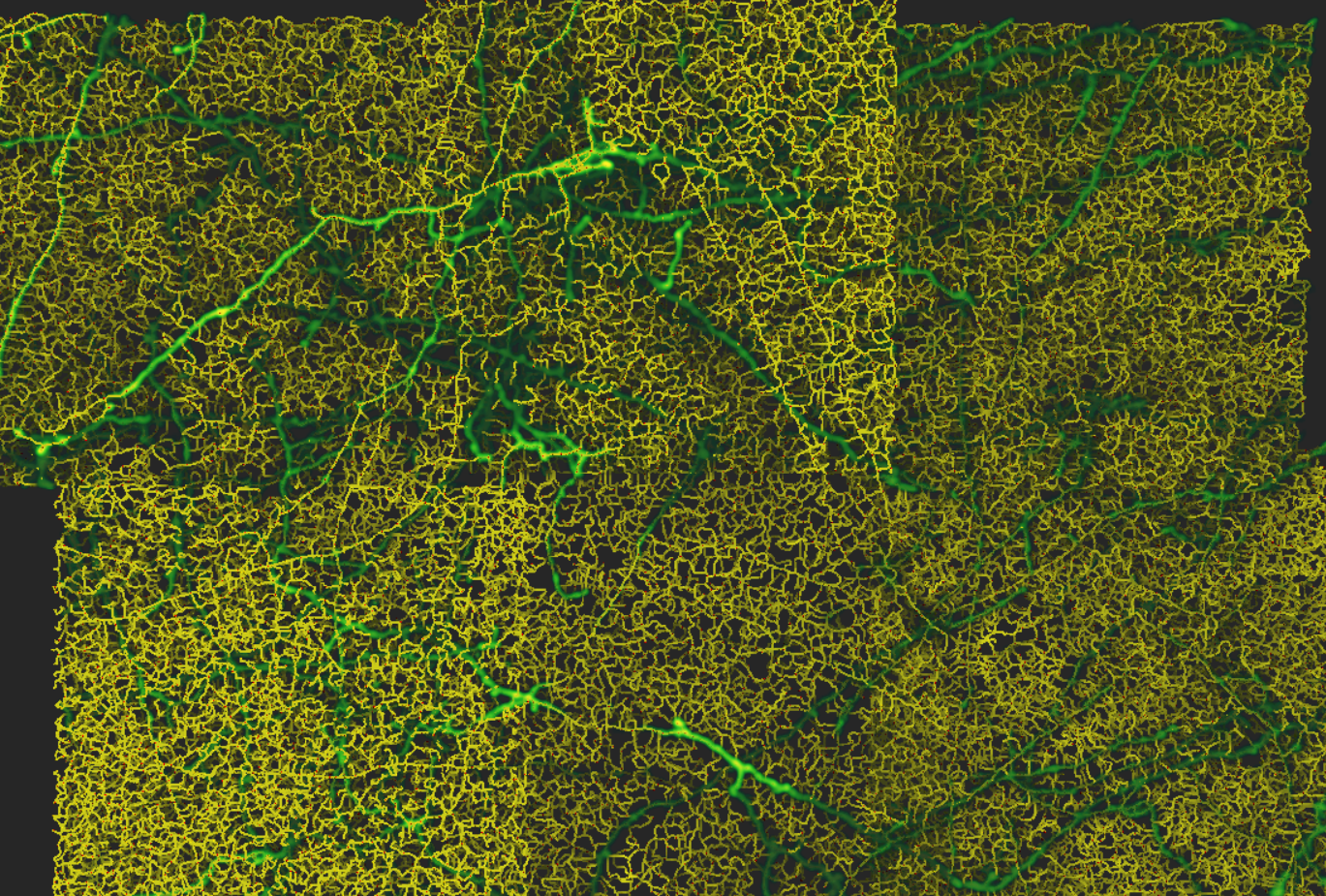}
		\vspace{-1.5em}
		\caption{\label{fig:lowpersrg}%
		Persistence threshold near 0.}
	\end{subfigure}
	\begin{subfigure}{0.48\columnwidth}
		\centering
		\includegraphics[width=\textwidth]{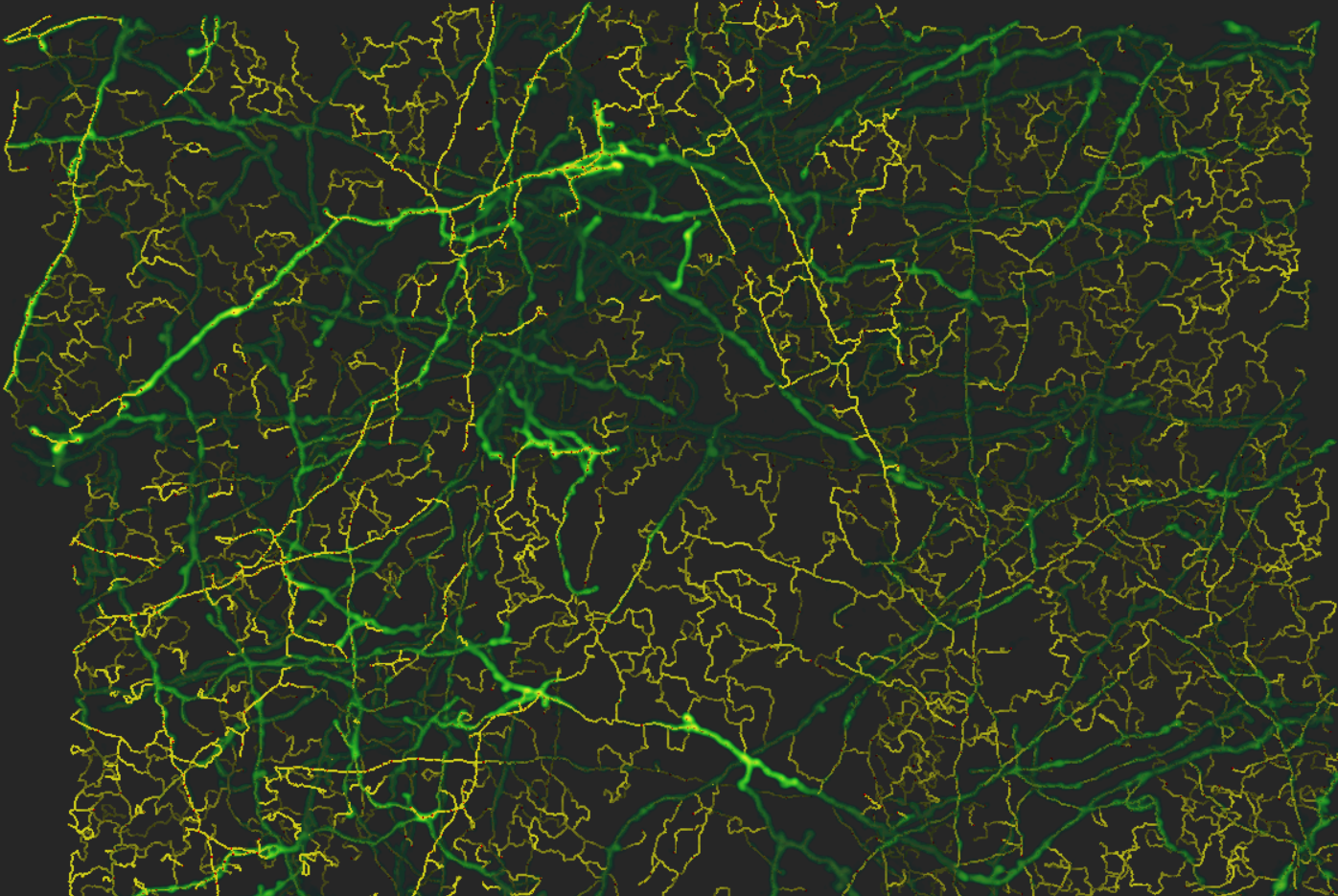}
		\vspace{-1.5em}
		\caption{\label{fig:highpersrg}%
		Persistence threshold at 1\%.}
	\end{subfigure}	
	\vspace{-1em}
	\caption{\label{fig:msc-all-arcs-desktop}%
	Low-threshold persistence simplification removes many extraneous features
	due to noise while preserving faint but desirable ones.}
	\vspace{-1em}
\end{figure}

\paragraph{MSC Computation and Simplification.}
We use a standard approach based on discrete Morse theory to construct a discrete gradient field~\cite{Robins11, Gyulassy2019}, available in the open-source MSCEER~\cite{MSCEER} library.
The library computes a discrete representation of the gradient using a parallel local filter,
after which it traces integral paths in the gradient field to construct the 1-skeleton of the MSC.
MSCEER also supports computing the MSC at a user-specified persistence
simplification threshold~\cite{Gyulassy2006},
which we use to simplify extraneous features created by noise.
Higher thresholds will produce coarser complexes and sparser
MSC-graphs;
however, selecting too high of a threshold may remove faint but desirable features.
We \revision{empirically }found that selecting a low threshold---as low as 1\% of the function range---is sufficient
to remove a large portion of the noise while keeping the majority of faint neurons
(\Cref{fig:msc-all-arcs-desktop}).
Finally, the 2-saddle-maximum arcs of the MSC are split as needed to remove geometric overlap, forming the MSC-graph.

\begin{figure}[t]
	\centering
	\begin{subfigure}{0.39\columnwidth}
		\centering
		\includegraphics[width=\textwidth]{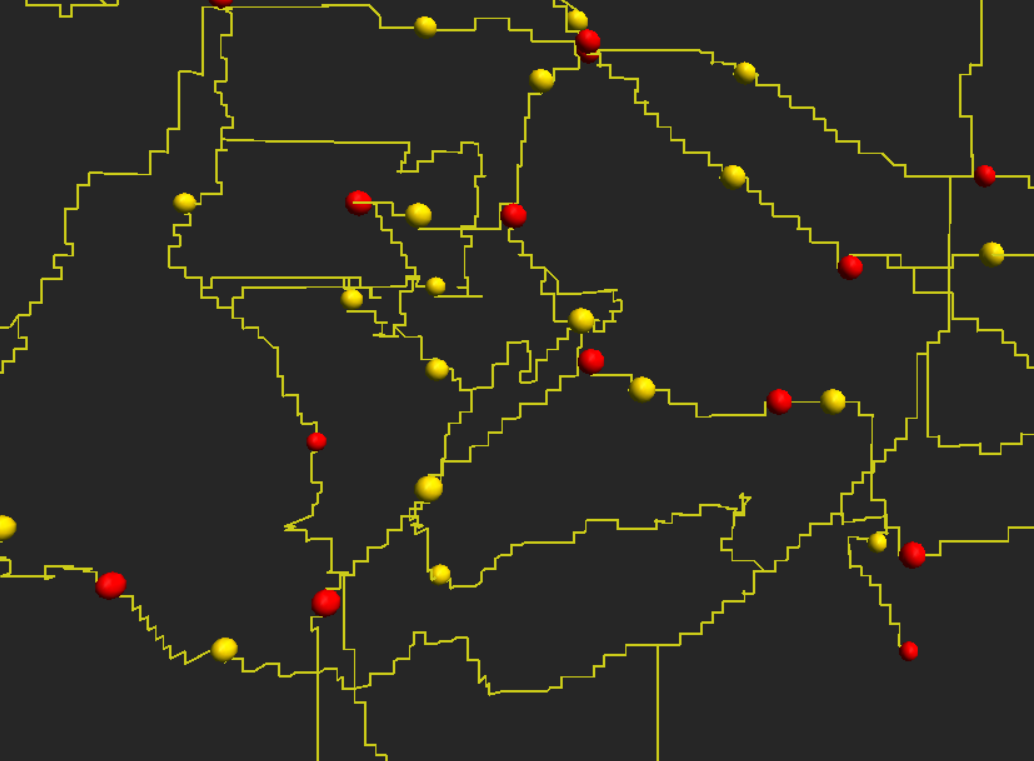}
	\end{subfigure}
	\begin{subfigure}{0.39\columnwidth}
		\centering
		\includegraphics[width=\textwidth]{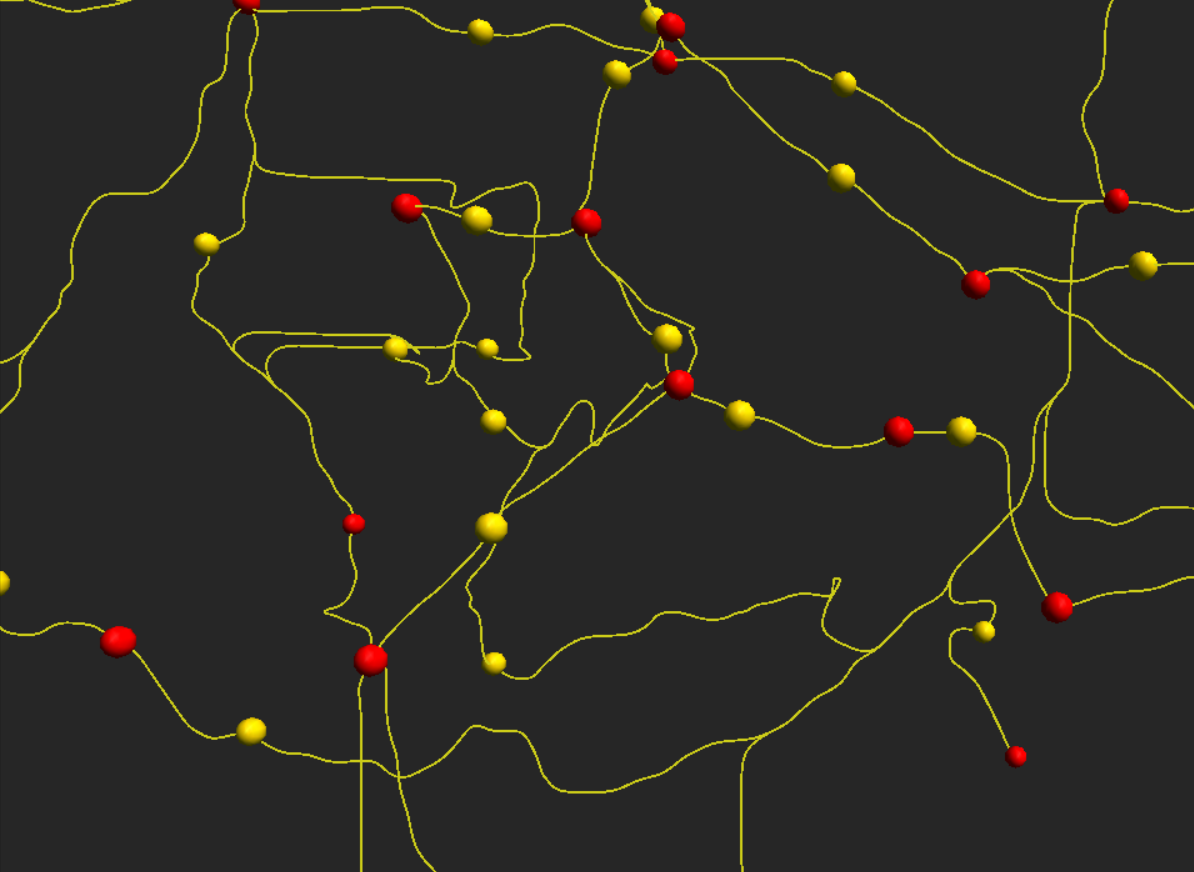}
	\end{subfigure}
	\vspace{-0.5em}
	\caption{\label{fig:smoothing}%
	The arcs and critical points from the original MSC-graph in its discretized form (left),
	and after applying the smoothing process (right).}
	\vspace{-1.5em}
\end{figure}

\paragraph{MSC-Graph Post-Processing.}
The discrete gradient used in computing the MSC and MSC-graph produces arcs whose
segments follow the staircase-like structure of voxels aligned with the underlying grid axes. These arcs are smoothed
using a simple averaging of neighbor positions to produce more
natural appearing results and improve arc length calculations (\Cref{fig:smoothing}).

 
\subsection{A Fast and Efficient Querying Framework}
\label{sec:msc-fast-query}

Our semi-automatic tracing method is based on adapting the MSC-graph into a data
structure that can quickly answer nearest neighbor and shortest path queries based on user input and navigation.
In VR, we require that all queries must return and update the rendering in well under 11ms to maintain 90~FPS, otherwise they can cause the visualization to skip frames, which is disorienting to users.
The key interactions in our system are picking the closest point in the MSC-graph,
picking all points within \revision{some radius},
and computing shortest paths between points in the MSC-graph.
The geometry of each arc of the MSC-graph is stored as an ordered set of 3D points. 
To accelerate nearest point and radius queries on the MSC-graph, we insert these points into a $k$-d tree, which is built at start up when loading the data.
Each point contains a reference to the arc in the MSC-graph from which it originated.
To find the nearest point to the controller, the $k$-d tree is queried and the corresponding arc looked up in the MSC-graph.

\revision{To find the best path reconstructing a neuron between some start and end point, we query the MSC-graph using
a weighted shortest path algorithm} (\Cref{fig:msc_tool_overview}c).
Paths corresponding to neurons have \revision{higher intensity values in the image}, thus we bias path selection towards paths \revision{through
voxels with higher data values}.
The weight of each arc is the integral of $w(p) = \epsilon + 1.0 - I(p)$ over the arc, where $I$ is the \textit{normalized} image intensity, $p$ is a point in the domain, and $\epsilon$ is a small constant to avoid zero-cost arcs.
The arc weights are computed at startup to improve \revision{runtime} performance.
When a selection query returns a point inside an existing arc, we symbolically split the arc at the point and recompute the weights of each of the new arc segments. 


To extract the shortest path in the graph, we use A*~\cite{hart_astar_1968}.
A* is similar to Dijkstra's algorithm, but prioritizes paths in the graph likely to
be the optimal path based on a heuristic, which
reduces the number of nodes in the graph that are processed in the search.
We select a standard heuristic that uses the L2 distance to the point being queried.


\section{Tracing Tool Design}
\label{sec:design}

Tracing neurons in complex, noisy, and poorly imaged \revision{regions often encountered in real-world data}
is not a straightforward task.
When tracing, the neuroscientist must make careful decisions to determine connectivity in \revision{such regions}.
Similarly, determining whether or not a neuron is branching \revision{is difficult in such regions, and those where} many
neurons \revision{intersect or pass in close proximity to each other.}
The design of our tool is motivated by two primary goals: to aid the user in better understanding the data
to facilitate the decision making process, and to increase the efficiency and ease of tracing neurons. 

Drawing from common best practices for design studies~\cite{sedlmair_design_2012}, we worked closely with domain
scientists, \revision{our co-authors}, to develop the tool through multiple iterations of testing and feedback.
Starting from an existing tool for manual neuron tracing~\cite{usher2018virtual}
(\Cref{sec:vrnt-framework}),
we integrated a prototype of our semi-automatic tracing method, which we refined based on feedback from domain scientists (\Cref{sec:msc-guided-vr}).

\subsection{Virtual Reality Neuron Tracing Framework}
\label{sec:vrnt-framework}
Our VR software framework is built using OpenVR, supporting \revision{room scale and seated modes}
using the variety of VR headsets available today, \revision{and is implemented in C++ and OpenGL.} In this work, we use an HTC Vive Pro.
The framework supports intuitive interaction modes for navigating volumetric data and manual neuron tracing,
along with streaming and rendering large volume
data through the IDX format~\cite{pascucci_global_2001} and an integrated data caching system.
We provide a brief overview of this framework, which served as the starting point for
implementing our MSC-guided tracing method, in the following two sections.


\subsubsection{Tracing and Navigation}
\label{sec:vrnt-framework-tracing}
Tracing neurons and navigating the data are the two primary 3D interactions performed when working
on a neuron reconstruction, and thus must be intuitive and quick to perform.
In the VR framework, one of the Vive controllers is mapped to tracing and the other to navigation.
Both interactions are initiated by holding the trigger on the respective controller and
moving it, to either trace along a neuron, or directly grab the volume and translate it.
Releasing the trigger then ends the interaction, either stopping the trace or releasing the volume.
The user can also navigate by walking in the virtual space.
To trace a neuron, the user moves the controller through space following the structure
as a line is drawn from the tip of the controller.
Navigating with the controller
is used to either stream new data from disk when working with large volumes, or to reduce the amount
of motion the user must perform, e.g., if working in a chair.

\revision{Capturing the branching structure of a neuron is critical
to reconstructing its connectivity, for use in subsequent analysis tasks.
Thus, it is important to support an intuitive workflow for tracing the branches of a neuron.}
The user can start a trace off any point on an existing one
to begin a new branch, or follow a branch back to its parent trace to connect it.
If mistakes are made during the tracing process, a quick undo operation can be performed by
pressing the trackpad. Corrections can be made when reviewing a trace by selecting portions
of the trace with the controller and deleting them by pressing the trackpad. The user can then re-trace
the removed section to correct it.
The tool uses haptic feedback to improve the user's perception of physically selecting
points on existing traces when creating branches or making edits by
playing a haptic click when hovering a point.





\subsubsection{Rendering}
\label{sec:vrnt-framework-rendering}
As scientists will potentially use the tool for hours on end to perform their work, providing a comfortable
experience is critical to avoid motion sickness or discomfort.
To meet the high-resolution and frame rate demands of VR, the
framework follows best practices from VR game development~\cite{Vlachos2015}.
The tool renders a $256^3$ subregion of the volume to keep the
rendering cost within an 11ms time budget to achieve 90FPS, and limits the amount of data paged
onto the GPU each frame. \revision{The data paging system allows exploration of arbitrary
sized volumes using the IDX format}~\cite{pascucci_global_2001}, \revision{combined with a CPU and GPU
on-demand data paging system that loads data onto the GPU as needed.}
The volume is stored in a sparse 3D texture, and
is rendered by a standard single-pass GLSL volume ray caster \revision{supporting
both volume and implicit isosurface raycasting} (e.g.,~\cite{stegmaier_simple_2005}).
\revision{Traces are rendered as thick lines using OpenGL line primitives.}
To further reduce the number
of pixels (and thus rays) that must be shaded each frame, the renderer
uses the \texttt{NV\_clip\_space\_w\_scaling} extension \revision{(available on NVIDIA GTX 10 series and newer GPUs)}
to reduce the rendering resolution at the edges of the eye, approximating a foveated
rendering approach. \revision{As users are typically focused at the center
of the image on the data being analyzed, the reduced resolution around
the edges of the image is not disruptive.}

\subsection{Semi-Automatic Tracing in Virtual Reality}
\label{sec:msc-guided-vr}

\begin{figure}
    \centering
    \includegraphics[width=0.95\columnwidth]{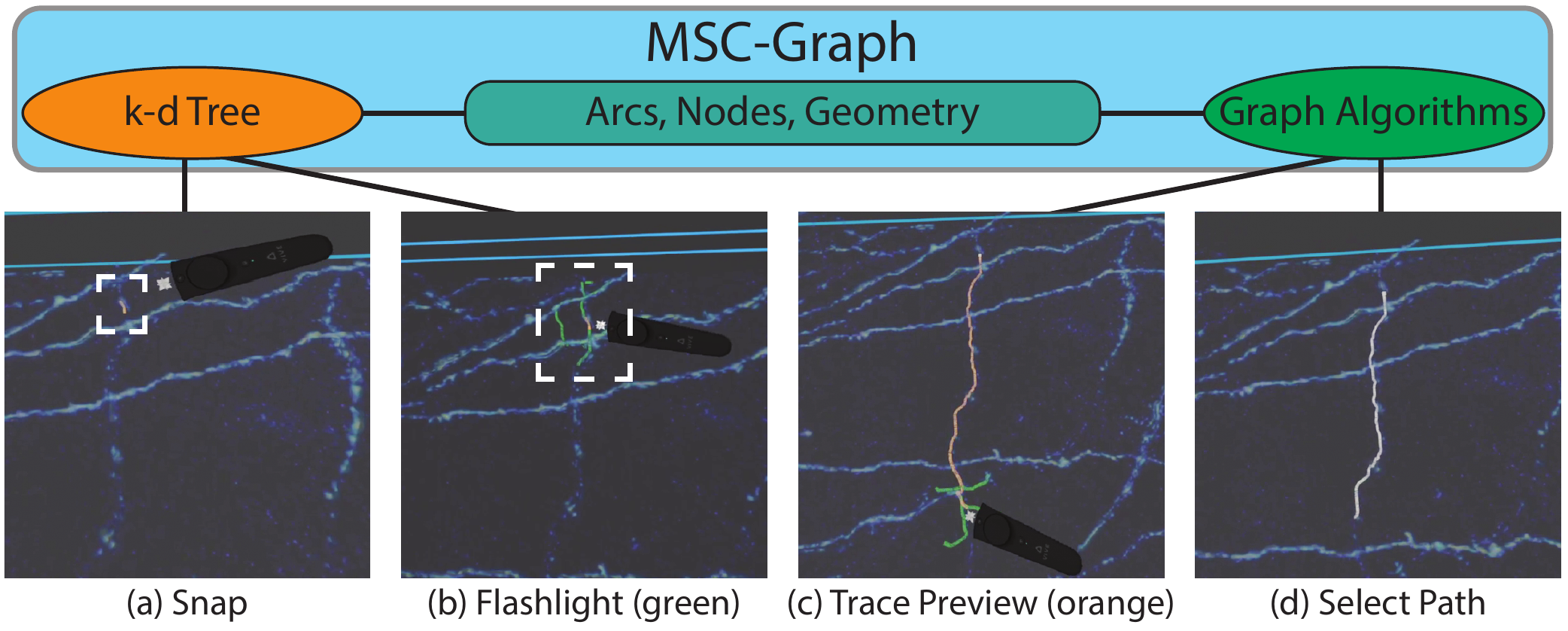}
    \vspace{-0.5em}
    \caption{\label{fig:msc_tool_overview}%
    The MSC-graph is leveraged to aid tracing in our VR system. 
    A trace interaction proceeds as follows:
    (a) When interacting, user selections are snapped to the center of the selected ridge.
    (b) The flashlight is used to preview local potential connections for guidance.
    (c) After picking a start point, the trace selected from the MSC-graph is displayed as a live preview
    while the user navigates to the end point.
    (d) Users can then accept the selection and add it to their reconstruction.}
    \vspace{-1.5em}
\end{figure}




\revision{A key focus in the design of our tool is on enabling users to work with the MSC-graph in an intuitive
manner, as neuroscientists are unlikely to be familiar with the underlying topological framework.}
Displaying the entire set of arcs
computed in the MSC-graph is distracting and overwhelming (\Cref{fig:lowpersrg}),
and may lead to following arcs in the MSC-graph that do not correspond to \revision{the desired} neurons.
\revision{In our design,} we put the data and the neuroscientist's interpretation of it first, and display
the arcs on-demand as the user hovers the tracing controller over regions of the data.
\revision{Through design feedback from the domain experts, we developed a ``flashlight'' that can be used to preview the potential neuron traces
in a region, leveraging the query system described in}~\Cref{sec:msc-fast-query} (\Cref{sec:msc-guided-augmenting}).
\revision{When tracing, we leverage the same query system to compute and display a preview
on the fly to the user, integrating the tracing and proofediting tasks into a single efficient workflow} (\Cref{sec:msc-guided-tracing}).
\revision{To allow quickly tracing through larger regions of the data we also integrated a zoom feature, which switches to a lower
resolution representation of the data to avoid frame drops.}
An overview of how the MSC-graph is leveraged in our tool to provide guidance to users
is shown in~\Cref{fig:msc_tool_overview}.

\subsubsection{Augmenting Visualization to Aid Users}
\label{sec:decision_making}
\label{sec:msc-guided-augmenting}

Our initial prototype only displayed the user's current selection (e.g.,~\Cref{fig:msc_tool_overview}c)
when tracing. Although this is valuable to display the selection so far, it does not provide
guidance on where to go next. To assist the user in choosing where to continue
the trace, \revision{especially at difficult decision points,} we provide an additional visual aid, the ``flashlight'' (\Cref{fig:msc_tool_overview}b).


The flashlight displays a preview of the nearby arcs in the MSC-graph \revision{as green lines}.
\revision{The arcs passing within a small ball around the controller are queried
from the $k$-d tree and line fragments outside the ball are discarded to avoid clutter.}
In \revision{noisy or poorly imaged} regions and ambiguous crossings
the flashlight allows users to peek into the underlying MSC-graph to gain
additional information about the potential connections to aid the decision making process.
For example, in regions with \revision{imaging} gaps, the flashlight can be used to check if the
underlying ridge line continues across the gap or not, and if it reconnects later to a region with
better imaging quality. 
Similarly, when deciding on crossings or branchings in regions \revision{with many nearby neurons}
the flashlight previews the set of possible paths which are most likely given
the topology of the data. By providing additional information that is less reliant on imaging quality and
visual representation, the flashlight is able to supplement the neuroscientist's
domain knowledge to \revision{make more informed decisions when tracing.}

\subsubsection{Tracing}
\label{sec:msc-guided-tracing}
When using the MSC-guided tracing tool, both the flashlight and the closest arc in the MSC-graph
are highlighted, giving a small live preview of the arcs in the underlying complex. The closest arc
is shown \revision{as an orange line} (\Cref{fig:msc_tool_overview}a,b), displaying the arc that will be selected when starting a trace.
To begin an MSC-guided trace, the user places the controller next to the
arc they want to start at and presses the trigger to pick a starting point within the arc, \revision{``snapping''
the selection to the arc.}
As they move the controller along the neuron being traced, \revision{we recompute the candidate trace using
the fast weighted shortest path algorithm on the MSC-graph discussed previously} to update the preview \revision{in real time}.
To end the trace and add the displayed preview to the \revision{reconstruction}, the user presses the trigger
at the desired end point (\Cref{fig:msc_tool_overview}c,d). The live preview
allows users to check that the selected path accurately follows the desired neuron
before selecting the trace.


\revision{When tracing manually, the user must hold} the trigger for the duration of the trace,
and trace each neuron segment individually and precisely.
However, when using our MSC-guided tool, they simply pick the start point and navigate to the
desired end point, while checking that the preview follows the desired structure.
The MSC-guided tool snaps the trace to the neuron's center line,
producing an accurate trace without requiring significant physical effort or precision from the user.
If the preview jumps away from the desired neuron, either because a shorter path is found or the MSC-graph lacks a needed connection,
the user places an end point before the problematic section and continues a new MSC-guided or manual trace off the end point.


To trace the branches of a neuron, the user can choose between 
a manual or MSC-guided trace to start from some point on the
existing tree, \revision{or start a new trace and reconnect it to the parent}.
When connecting a new MSC-guided branch \revision{to an} existing parent tree,
it is possible that no arc exists in the MSC-graph to connect the two. In this case, we create a connection to join the new branch
with the nearest point on the parent tree, if the gap is less than a few voxels. 

When evaluating our first prototype, the neuroscientists found \revision{the MSC-guided tracing mode}
especially useful for tracing long axons through the data.
The ability to simply place the start point and navigate to the end point reduced the amount of physical
effort required, allowing them to focus on interpreting the data rather than precise interaction.
\revision{However, to navigate to the end point users would spend a large amount of time
translating the volume to find the end of the axon, limiting the speed at which they could
trace.}

\revision{To allow navigating at larger scales or getting an overview or close up view of the
data, we added a zoom interaction, performed by holding the grip buttons on both controllers and moving
them further apart or closer together.
Zooming out increases the number of voxels visible in the focus region, potentially degrading
performance below 90~FPS. To maintain an acceptable framerate, we leverage
the support for multiresolution queries provided by the IDX format}~\cite{pascucci_global_2001}.
\revision{In addition to the full resolution volume cache, we run a second lower resolution one
in parallel. When zoomed out, we switch to the lower resolution data to reduce rendering load;
when the user returns to the original zoom level we switch back to the full resolution data.
The caches are run simultaneously to ensure that the required data is available to provide
a smooth transition between levels when zooming in or out.}
\section{Evaluation}
\label{sec:evaluation}
To evaluate the effectiveness of our MSC-guided semi-automatic
neuron tracing method we study both the effectiveness of the underlying
topological framework, and the design of our semi-automatic tracing tool in VR\@.
First, to demonstrate that the MSC-graph provides an effective
framework for neuron tracing
we perform an offline comparison against semi-automatic
methods available in current desktop software, solely comparing the path-finding capabilities of each approach (\Cref{sec:offline-comparison}).
We then evaluate our MSC-guided tracing tool in virtual reality through
a pilot study with trained neuroanatomists and
undergraduate students (\Cref{sec:expert-eval}).

\begin{figure}
	\centering
	\includegraphics[width=0.9\columnwidth]{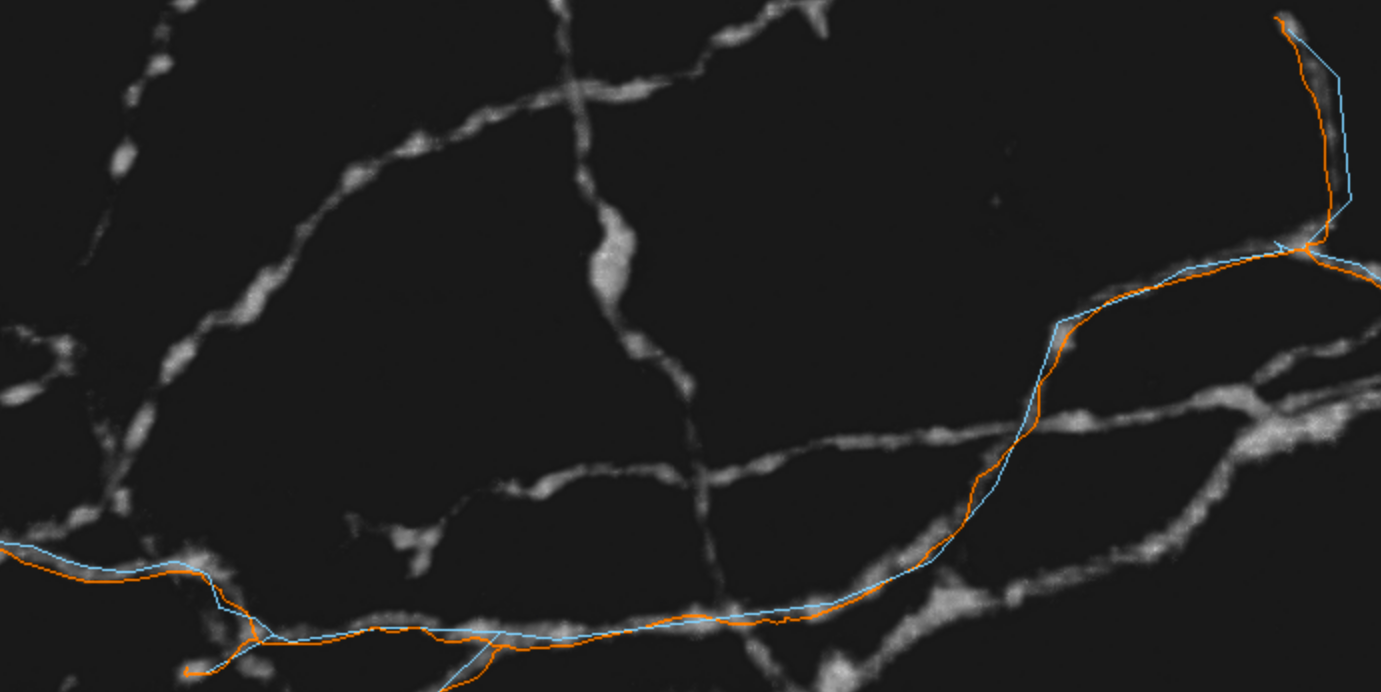}
	\vspace{-0.5em}
	\caption{\label{fig:diadem-vs-vr-expert}%
	The DIADEM reference (blue) was made using standard desktop software, and consists of
	coarse line segments. The VR reference (orange) consists of finer
	segments, and follows the neuron more closely.}
	\vspace{-1em}
\end{figure}

\subsection{Experimental Setup}

We evaluate our approach using two data sets.
The first data set is the \textit{Neocortical Layer 1 Axons} data set~\cite{depaola_nclayer1_2006}, made
publicly available for the DIADEM challenge~\cite{brown2011diadem}.
The data set is a $1464\times1033\times76$ volume made from six aligned subvolumes containing 34 axons imaged
from a mouse brain. The resolution of the data is $\approx 0.08 \mu m$/pixel in X and Y and
$\approx 1 \mu m$/pixel along Z.
The data set includes a reference trace for each neuron, which we use as one point
of comparison in our evaluation. The reference traces were used in the DIADEM challenge
and produced manually using NeuroLucida.
Throughout the text we will refer to these traces as
the ``DIADEM reference traces''. 
We also compare against traces created manually by an expert in VR during a previous study~\cite{usher2018virtual},
referred to as the ``VR reference traces''.
As neuron traces are produced by hand by experts, there is an inherent subjectivity in each
trace informed by the expert's knowledge of the data and imaging process,
and as such no true ``ground truth'' is available to compare against.
During a review of the DIADEM traces we observed that they would often drift from the neuron, following
a straighter path than the underlying data, while the VR reference traces
followed the structure more closely (\Cref{fig:diadem-vs-vr-expert}).

The second data set, \textit{Cell Bodies}, was provided by 
A. A.'s laboratory~\cite{cellbodies} and is a 
$1024\times1024\times314$ volume with a resolution of $0.331\mu m$/pixel in X
and Y and $1.5 \mu m$/pixel in Z. The volume was imaged from a Marmoset visual
cortex and contains multiple cell bodies with axons that have a 
complex branching structure. The data set contains significantly more noise, \revision{overlapping neurons}, and poorly
imaged regions than the \textit{Neocortical Layer 1 Axons} data set, making it more difficult to trace.
The data set does not include a set of reference traces, in our comparisons we used the most experienced tracer's (subject 1) manual trace as the reference.

\begin{figure}[t]
    \centering
    \begin{subfigure}{0.32\columnwidth}
        \centering
        \includegraphics[width=0.5\textwidth]{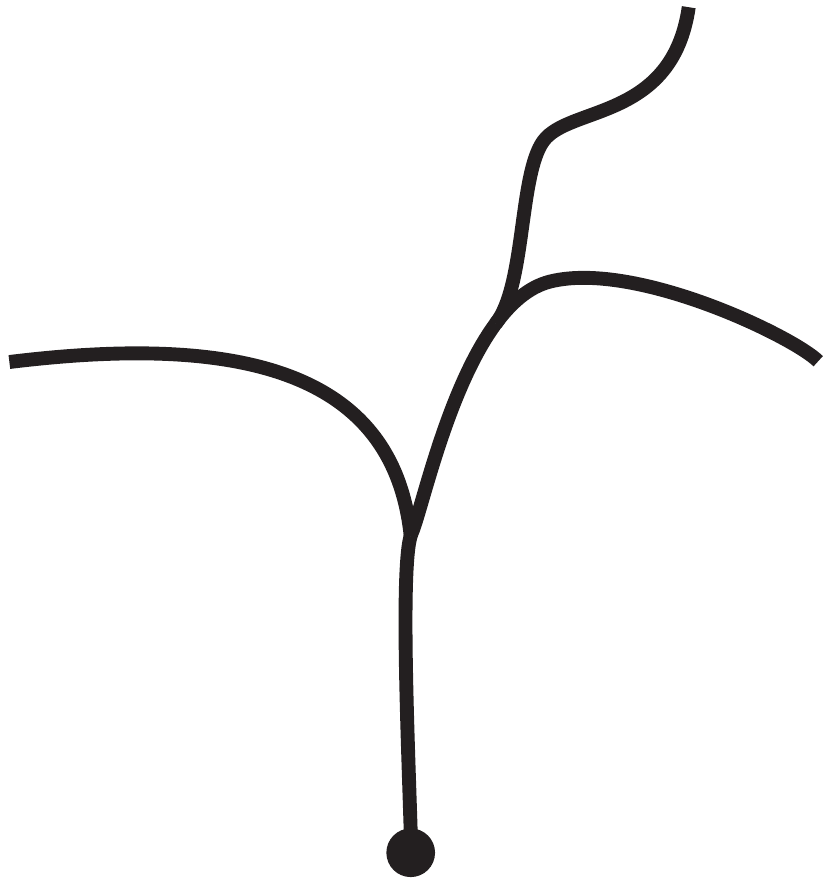}
        \caption{\revision{Reference.}}
    \end{subfigure}
    \begin{subfigure}{0.32\columnwidth}
        \centering
        \includegraphics[width=0.5\textwidth]{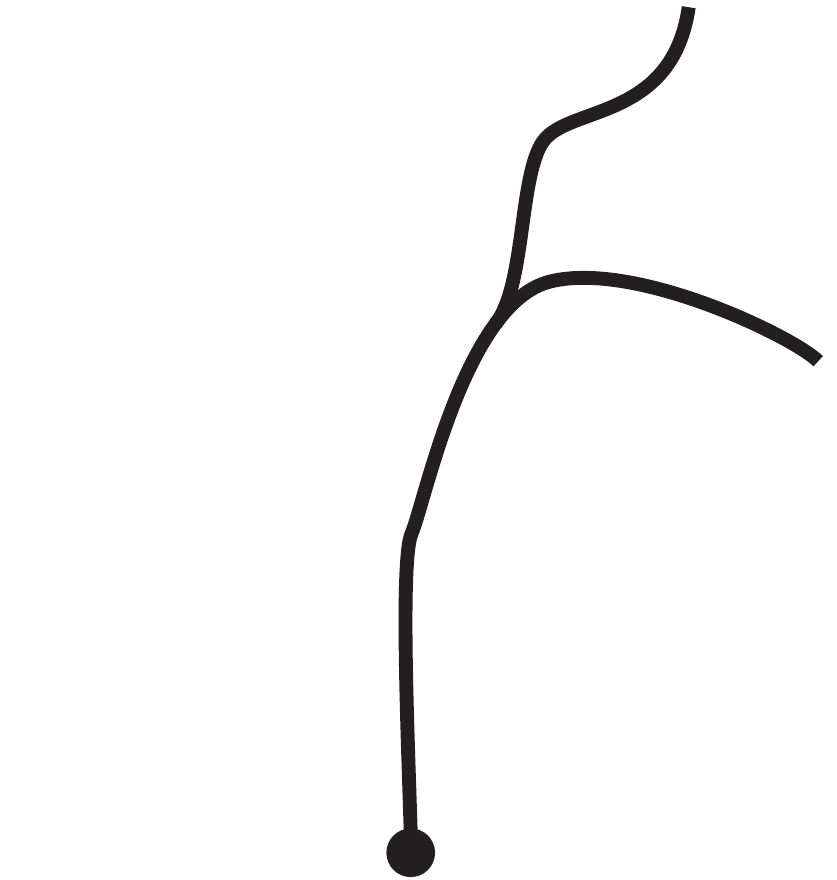}
        \caption{\revision{Branch missing.}}
    \end{subfigure}
    \begin{subfigure}{0.32\columnwidth}
        \centering
        \includegraphics[width=0.5\textwidth]{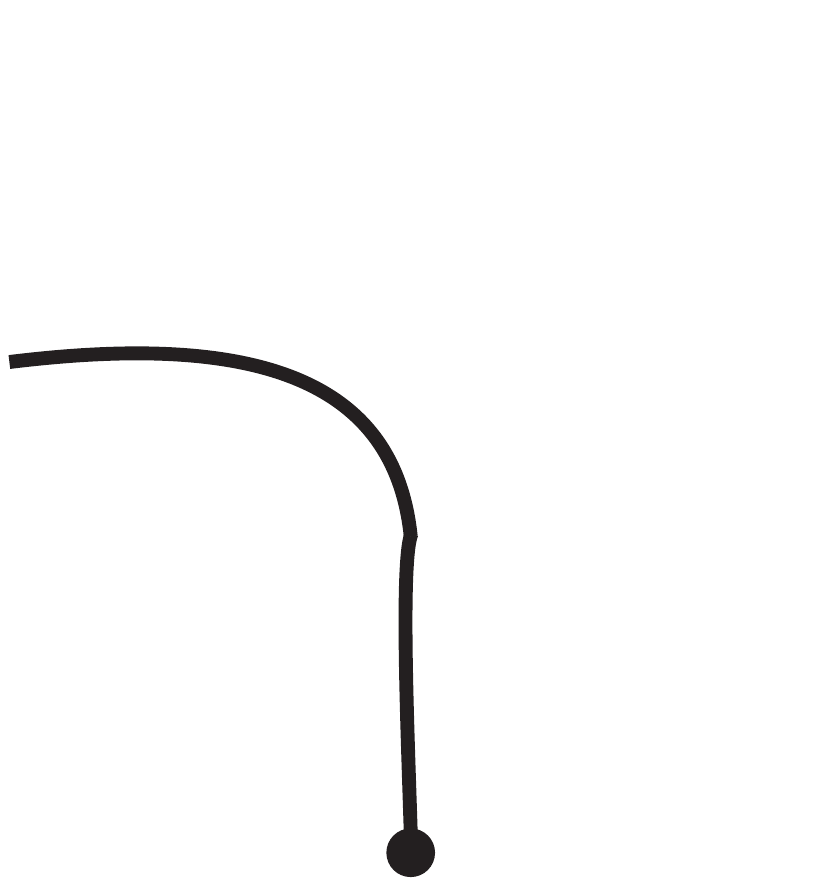}
        \caption{\revision{Subtree missing.}}
    \end{subfigure}
    \vspace{-0.5em}
    \caption{\label{fig:trace_score_example}%
    \revision{Examples of incorrect traces (b,c) compared to a reference trace (a).
    Traces (b) and (c) score 0.875 and 0.5 with the DIADEM metric, respectively.
    The error in (c) misses a subtree containing more important features
    than are missed in (b), impacting subsequent analysis significantly,
    and is scored lower as a result.}\vspace{-2em}}
\end{figure}


\paragraph{Scoring Trace Quality.}
We use the DIADEM metric~\cite{gillette_diadem_2011} to compare traces against the references and expert judgment to assess their accuracy and quality.
The DIADEM metric accounts for the \revision{length and topology}
of a trace, scoring how well a trace captures the branching structure of a neuron on a scale of 0 (dissimilar) to 1 (identical).
The score is penalized for missing branches, excess branches, incorrectly placed branches, and differences
in branch length (\revision{see}~\Cref{fig:trace_score_example}).
Branches and endpoints are matched to the reference by checking within a fixed radius of the reference points. 
The DIADEM score correlates reasonably well with expert judgment of trace quality
\revision{and remains widely used in the field}~\cite{magliaro_gotta_2019}.
However, \revision{the score focuses on the topology of the trace} and does not account for geometric differences.
Furthermore, the traces produced in our tool are more finely sampled along the trace,
which we have found to pose difficulties for the DIADEM scoring method
\revision{as the length of the trace will be longer than a similar trace composed of coarser line segments} (see \Cref{fig:diadem-vs-vr-expert}).
Developing metrics for robustly comparing reconstructions of the same neuron
remains an open problem~\cite{peng_bigneuron_2015,mayerich_netmets_2012}.
Thus, we also evaluate traces using expert judgment, and have one of the expert neuroanatomists rate
the quality of an anonymized subset of traces.
The expert rates the trace on how well it follows the center-line of a neuron and how accurately branches and
end points have been traced.

\subsection{Algorithm Comparison}
\label{sec:offline-comparison}
We first perform an offline study to assess the quality of the neurons computed
by our MSC-graph based algorithm compared to Vaa3D's semi-automatic tracing method~\cite{peng2010automatic,peng_v3d_2010}.
Vaa3D is a widely used open-source software suite for neuron reconstruction,
and is the designated platform for testing algorithms in the Big Neuron Project~\cite{peng_bigneuron_2015}.
Vaa3D's semi-automatic tracing works similar to our MSC-guided tool:
given a start and end point it will attempt to trace the neuron between them. 



\begin{table}
	\centering
	\begin{tabular}{@{}lrr@{}}
		\toprule
		Reference Trace & Vaa3D & MSC-guided \\ 
		\midrule
		DIADEM & 0.45 $\pm$ 0.37 & 0.74 $\pm$ 0.26 \\ 
		VR     & 0.53 $\pm$ 0.38 & 0.81 $\pm$ 0.22 \\ 
		\bottomrule
	\end{tabular}
	\vspace{-0.5em}
	\caption{\label{table:offline-diadem-score}%
	DIADEM scores for traces extracted with Vaa3D and our MSC-guided method
	\revision{compared to} the reference traces.
	We find that the MSC-guided method computes better and more consistent traces.}
	\vspace{-1em}
\end{table}

\begin{table}	
	\centering
	\begin{tabular}{@{}lrr@{}}
		\toprule
		  User & Manual Score & MSC-guided Score \\
		\midrule
		\multicolumn{3}{c}{Compared to DIADEM Reference Traces} \\
		\cmidrule{1-3}
		\stwo   & 0.51 $\pm$ 0.35 & 0.43 $\pm$ 0.36 \\
		\sfour  & 0.45 $\pm$ 0.37 & 0.35 $\pm$ 0.35\\
		\seight & 0.35 $\pm$ 0.35& 0.25 $\pm$ 0.31 \\
		\snine  & 0.26 $\pm$ 0.35 & 0.23 $\pm$ 0.28\\
		\sten   & 0.42 $\pm$ 0.34 & 0.32 $\pm$ 0.31\\
		\midrule
		\multicolumn{3}{c}{Compared to Expert VR Traces} \\
		\cmidrule{1-3}
		\stwo   & 0.46 $\pm$ 0.46  & 0.40 $\pm$ 0.41\\
		\sfour  & 0.56 $\pm$ 0.45 & 0.28 $\pm$ 0.39 \\
		\seight & 0.36 $\pm$ 0.42 & 0.24 $\pm$ 0.38 \\
		\snine  & 0.34 $\pm$ 0.42 & 0.30 $\pm$ 0.37 \\
		\sten   & 0.44 $\pm$ 0.41 & 0.33 $\pm$ 0.38 \\
		\bottomrule
	\end{tabular}
	  \vspace{-0.5em}
	  \caption{\label{table:diadem-scores}\label{table:vr-ref-scores}%
	  DIADEM scores \revision{ (average and standard deviation) comparing users'}  manual and MSC-guided traces \revision to {DIADEM reference traces (top), and an expert's manual trace in VR (bottom)}.}
	  \vspace{-1.5em}
\end{table}
\begin{table*}	
  \vspace{-0.5em}
  \centering
  \relsize{-1.5}{
\begin{tabular}{l|rrrrrr|rrrrrr|rrr}
\toprule
     & \multicolumn{6}{c}{Session 1}                                & \multicolumn{6}{c}{Session 2}                                & \multicolumn{2}{c}{Total}                                    & \multicolumn{1}{c}{}   \\
     \midrule
     & \multicolumn{3}{c}{Manual: Set A} & \multicolumn{3}{c}{MSC\_guided: Set B} & \multicolumn{3}{c}{Manual: Set B} & \multicolumn{3}{c}{MSC\_guided: Set A} & \multicolumn{1}{c}{Manual} & \multicolumn{1}{c}{MSC\_guided} & \multicolumn{1}{c}{}   \\
     \midrule
User & Time  & Dist   & Rate & Time    & Dist    & Rate   & Time  & Dist   & Rate & Time    & Dist    & Rate   & Rate                  & Rate                       & Speedup                \\
1    & 3275  & 34453  & 10.5      & 2532    & 23809   & 9.40        & 2337  & 23963  & 10.25     & 2401    & 29804   & 12.41       & 11.06                      & 10.48                           & \textbf{0.94$\times$} \\
2    & 1403  & 19911  & 14.19     & 982     & 19751   & 20.11       & 1267  & 18883  & 14.90     & 1251    & 25257   & 20.18       & 14.12                      & 19.91                           & \textbf{1.41$\times$}  \\
3    & 1979  & 39555  & 19.98     & 1525    & 41329   & 27.10       & 2369  & 38820  & 16.38     & 1957    & 35237   & 18.00       & 17.63                      & 22.5                            & \textbf{1.27$\times$}  \\
4    & 1823  & 9844   & 5.39      & 2142    & 17391   & 8.11        & 2345  & 11539  & 4.92      & 1524    & 10397   & 6.82        & 5.14                       & 7.57                            & \textbf{1.47$\times$}  \\
5    & 1334  & 17443  & 13.07     & 1350    & 37825   & 28.01       & 2366  & 20532  & 8.67      & 1406    & 41906   & 29.80       & 9.95                       & 27.3                            & \textbf{2.74$\times$}  \\
\bottomrule
\end{tabular}
}
	  \vspace{-0.5em}
      \caption{\label{table:tracing-times}%
	  Speedups for total task time of the MSC-tool vs.\ manual tracing. Time is in seconds, distance is in voxels, and rate is the distance traced over the time, the total across each task set is reported. Overall, the improvement in tracing times was higher using the MSC-guided method. }
        \vspace{-1.75em}
\end{table*}

The guide points are generated from the existing DIADEM and expert VR traces by extracting
the start, branch, and end points to simulate user clicks.
The points are extracted automatically through a depth-first traversal of the neuron tree, and supplied in that order.
This approximates how a user would interact with
a semi-automatic method, clicking along the structure to mark key points
on the neuron and letting the algorithm extract the structure.
Our MSC-guided method works between a
single start and end point, thus both start and end points are provided for each segment of the neuron.
Along with point to point extraction, Vaa3D can
trace from a start point to connect to a set of points, which is the mode we use in our evaluation.
In both cases, the paths between the points are automatically generated. 



We found that the traces computed by our MSC-guided method
follow the neuron more accurately and achieved a higher DIADEM score than those computed using Vaa3D's semi-automatic method (\Cref{table:offline-diadem-score}).
This can be partly
attributed to the MSC-guided method placing greater weight on
the user's guidance, extracting the shortest path between
the given end points as they are clicked in order.
Vaa3D's semi-automatic method treats the set of points as
hints and is not guaranteed to connect them in the same order.

Our MSC-guided approach is able to provide faster neuron extraction times,
as required for VR.
Vaa3D took 28.64s per reconstruction on average, while
our MSC-guided method took just 0.029s, a speedup of $986\times$.
Our MSC-guided method computes shortest paths on the relatively sparse MSC-graph (\textasciitilde10k nodes) compared to the implied
adjacency graph of the original voxel data (\textasciitilde100M voxels). 
The precomputation to build the MSC-graph is fast
and scalable~\cite{MSCEER,Gyulassy2019};
on a laptop with an i7--7700HQ CPU
the image filtering and blurring takes 26s using ImageJ,
after which the MSC-graph is computed in 134s using MSCEER~\cite{MSCEER}. Although the total time is longer
\revision{for a single neuron, the MSC-graph can be reused for all neurons in the volume, allowing real-time path computation in VR.}

\begin{figure}
	\centering
	\includegraphics[width=0.95\columnwidth]{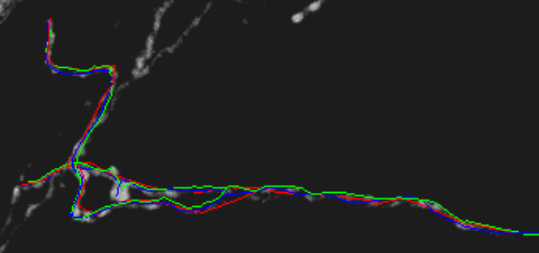}
	\vspace{-0.5em}
	\caption{\label{fig:visual-msc-vs-diadem-vr}%
	An MSC-guided trace (green) has lower DIADEM score than a manual (blue) VR trace (0.529 vs. 0.833, respectively), when compared against the corresponding DIADEM (red) reference trace, even through the manual trace contains an extra branch.
	\revision{Both the MSC-guided and manual traces were rated acceptable by an expert}.
	It is possible that differences in the length of segments, or failure to match branch points between MSC-guided and DIADEM traces contribute to a low score.}
	\vspace{-1em}
\end{figure}

\subsection{User Study}
\label{sec:expert-eval}
We tested our MSC-guided software infrastructure through a pilot study, where neuroscientists performed real traces on a varied set of tasks.
We evaluate our tool using quantitative measures of trace quality and speed, and extensive qualitative feedback from the user group
on the overall impact on workflow and user experience.


\begin{figure}[t]
	\centering
	 \begin{subfigure}{0.49\columnwidth}
		 \centering
		 \scalebox{1}[-1]{\includegraphics[width=\textwidth]{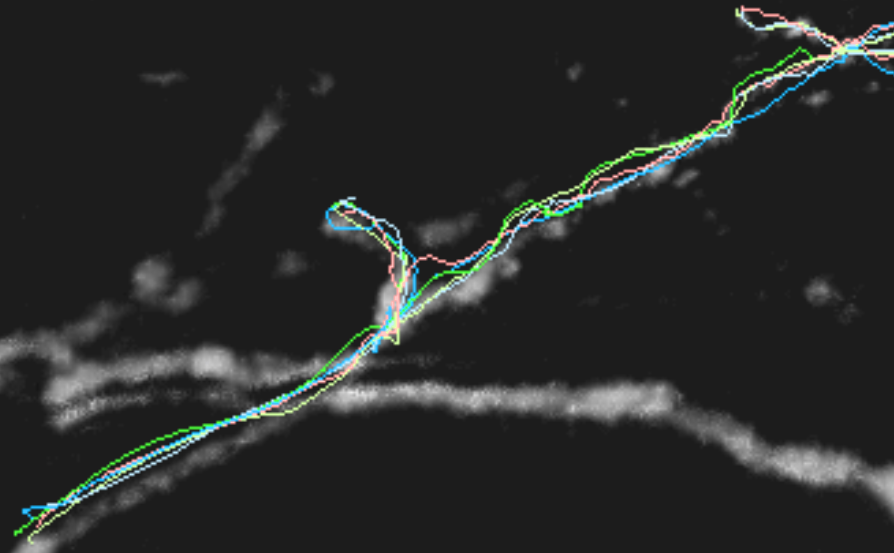}}
		 \vspace{-0.5em}
		 \caption{\label{fig:man_alltogether}%
		 Manual traces.}
	 \end{subfigure}
	\begin{subfigure}{0.49\columnwidth}
		\centering
		\scalebox{1}[-1]{\includegraphics[width=\textwidth]{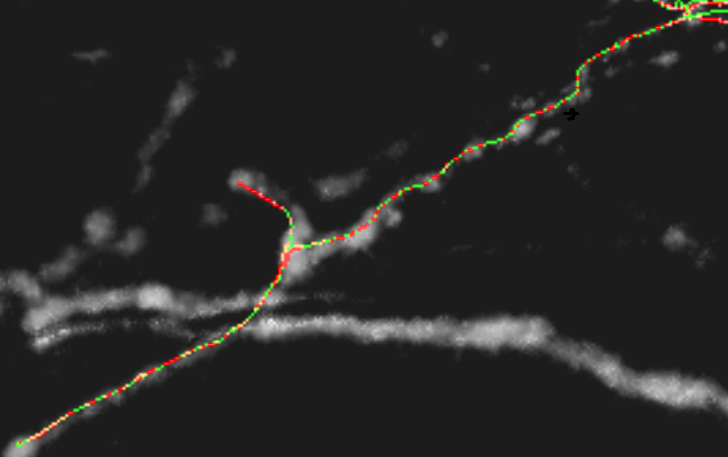}}
		\vspace{-0.5em}
		\caption{\label{fig:mscalltogether}%
		MSC-guided traces.}
	\end{subfigure}
	\vspace{-1em}
	\caption{\label{fig:all_trace_comparison}%
	A comparison of the same trace from all users. The left shows how manual traces can be inconsistent, while the right hows the high consistency using the MSC-guided method.}
	\vspace{-1.5em}
\end{figure}

\begin{figure}[t]
	\centering
	\includegraphics[width=0.9\columnwidth]{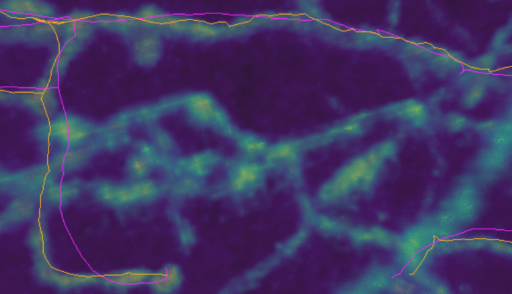}
	\vspace{-0.75em}
	\caption{\label{fig:man_deviation}%
	Users tracing manually (pink) can significantly deviate from the neuron. The MSC-guided trace (orange), automatically snaps to the center of the neuron, producing a more accurate trace. \revision{These traces were produced by subject 3.}}
	\vspace{-1.5em}
\end{figure}

\paragraph{User group.}
We conducted our pilot study with five users of varying levels of experience from A. A.'s laboratory. 
Subjects 1 and 5 are senior neuroanatomists,
subjects 2 and 3 are undergraduates with 2--3 years
of experience tracing neurons using NeuroLucida, and
subject 4 is a graduate student  with little prior tracing experience 
but with advanced knowledge of neuron morphology.
This range of experience levels provides a representative sample of
a typical connectomics lab. 

\paragraph{Description of tasks.}
Each subject traces a set of 34 neurons twice over two separate sessions,
spaced at least three days apart \revision{to ensure users did not learn specific traces from a previous session}.
The first 28 neurons come from the
\textit{Neocortical Layer 1 Axons} data set, the last six are from the \textit{Cell Bodies} data set.
For each neuron the start point is marked in space and the user instructed
to trace the neuron to its perceived end points.
In each session \revision{the users traced the first} half of the neurons manually
and \revision{the second} half using any desired combination of the MSC-guided tool and manual tracing.
In the second session the set of neurons which were traced manually and with the MSC tool is flipped.
At the start of the their first session, each user completed a training session
to learn the manual and MSC-guided tools.

\paragraph{User training.}
The goal of the training was to reduce the effect of learning on evaluating the accuracy and speed of manual and MSC-guided tracing.
The training process acquainted users with the VR system, and demonstrated the capabilities and limitations of our MSC-guided tracing, and how to address them.
The user was given four starting points in one of the test data sets and asked to trace the neurons to completion. When they finished tracing a neuron, they were shown their trace compared to a reference.
To emphasize how their trace may have deviated from the reference, their trace was colormapped
by the shortest distance to a corresponding point on the reference.
\revision{By the end of the training, users demonstrated an intuitive feel for how the MSC-guided tool operated.}
In particular, the senior neuroanatomist recommended the training process not only for familiarizing users with the tool, but also to teach novices the reasoning behind selecting paths, branch points, and start and end points.

\paragraph{Interactive session.} 
Each of the sessions was held in a room with enough space to walk around and explore the data, using a 2.5m$\times$2m VR area, under supervision of the authors. While the users completed each task, their interactions, i.e., tracing, panning, zooming, head and hand movement, and controller button clicks, were recorded by the tool and exported to a JSON file for analysis. On average, each session took one to one-and-a-half hours.
\revision{Users were able to walk freely or sit as desired for any portion of the session.}
\subsubsection{Quantitative Evaluation of Traces}
\label{sec:user_trace_quant_eval}
Tracing neurons is a challenging task and essentially relies on the expertise and accuracy of the user.
This makes the task subjective, and can lead to disagreement on so-called ``reference traces'' produced by other experts have produced.
Hence in this scenario, we have to consider that scoring the user traces against a fixed reference does not necessarily mean the reference is more accurate. 

When using the DIADEM metric, we find that the quality of the traces produced using fully
manual tracing and our MSC-guided semi-automatic tracing are
similar, with traces made using the MSC tool receiving slightly lower DIADEM
scores (\Cref{table:diadem-scores}). Note that \textit{even compared to a reference manual VR trace}, DIADEM scores for traces also completed manually in VR were on average below 0.4. This indicates that either the DIADEM metric is very sensitive, or that the tracing task is so challenging that even experts disagree and produce different traces. Finally, the DIADEM score, while a reasonable metric, was
found to produce unexpectedly low scores on traces rated by our two experts
as acceptable.


\begin{figure}
	\centering
	\vspace{-0.5em}
	\includegraphics[width=0.85\columnwidth]{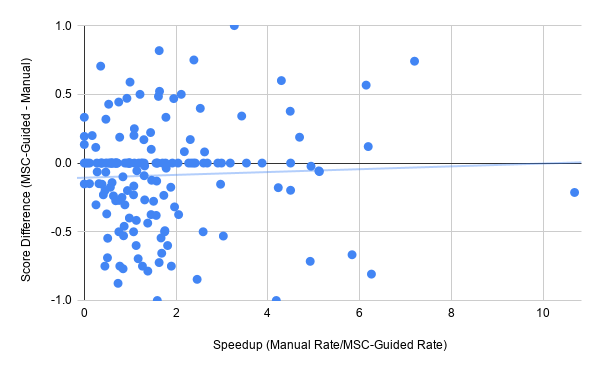}
	\vspace{-1.25em}
	\caption{\label{fig:correlation}%
    The correlation between score difference (MSC-guided score- manual score), and the speedup (Manual Rate/MSC-guided rate) of all individual tracing tasks. There is no correlation between the score difference and the speedup.}
	\vspace{-1.75em}
\end{figure}

To perform an additional validation of the traces we asked one of the senior neuroanatomists
in A.A.'s lab to review a subset of them. 
The expert reviewed a series of traces, each containing three anonymized traces of the same neuron:
one traced using the manual method, one using the semi-automatic method, and one being the DIADEM reference.
The expert noted that, due to the ambiguity in the data, most of the traces where potentially correct.
However, for almost every set reviewed, the expert rated the MSC-guided traces as more accurate
than those performed manually (\revision{see}~Figure~\ref{fig:visual-msc-vs-diadem-vr}).

In reviewing the user traces, we found a higher degree of geometric inconsistencies in the manual traces compared to the MSC-guided traces (see~Figure~\ref{fig:all_trace_comparison}).
Figure~\ref{fig:man_alltogether} shows how the hand motions of tracing neurons manually can significantly vary, while MSC-guided tracing (\Cref{fig:mscalltogether}) snaps the traces to the centerline of the neuron.
\revision{The MSC-guided traces have more consistent and higher geometric accuracy as a result.}
Figure~\ref{fig:man_deviation} highlights how geometrically inaccurate manual traces can be. If a user is not precise in their hand movements, they can produce traces that clearly deviate from the intended path.
We found this to frequently occur in places where the neurons bend either slightly or significantly.

Table~\ref{table:tracing-times} shows the total session time, distance traced, and the \revision{rate of tracing} in voxels per second for each set of tasks. We found that for the majority of users, the MSC-guided approach reduced the time it took to complete each tracing task compared to manual tracing. We explored whether there was a correlation between the speedup of a tracing task and the score of that task (\Cref{fig:correlation}).
The Pearson correlation coefficient between the two was 0.044, indicating no association between speedup and score, i.e., equivalent quality traces were produced in less time.
The users did not show consistent speedup or slowdown between the first and second sessions, indicating that memory of traces that have already been seen in the first session did not impact tracing in the second session.
We note that user 1 was slower when using the MSC-guided method, as he spent additional time exploring the data with the flashlight to find connections in sparse regions below the visible range of the volume or isosurface visualization. 

We remark on the significant variability in the total distance traced between users, both within and between manual and MSC-guided tasks. Empirically, this variability highlights the subjectivity in neuron tracing, where some users decide that neurons in certain complex or poorly imaged areas either branch or continue, while others decide they terminate.
We did not find a consistent increase or decrease in the total distance of neurons traced between MSC-guided and manual methods.  

The median speedup across all users was 1.41$\times$, with subject 5, an expert in their field, achieving the highest speedup of 2.74$\times$. Neuron tracing is a time consuming process, with typical real world traces taking days, weeks, or months. Reducing the time it takes to trace neurons by these factors could save hours or days of work, significantly accelerating the neuron reconstruction pipeline.

\subsubsection{Qualitative Expert Evaluation and Feedback}
\label{sec:expert-discussion}

In this section we detail our users' qualitative feedback regarding the
design and usability of the MSC-guided tracing tool, discussing both
benefits and limitations.
During the study we collected feedback from users
through a survey completed after each session and open-ended discussions.
The survey focused on the usability and usefulness of the MSC-guided
tool for tracing neurons, with questions rated on a 5-point Likert scale.
Open-ended discussions were used to solicit general feedback on the design of the tool
and general comments or issues regarding its use in practice.


\paragraph{User Experience.}
Overall feedback from users on the MSC-guided tool was positive.
In the survey all subjects reported preferring the MSC tool over
manual tracing, finding it less fatiguing \revision{and more comfortable to use for long periods}.
In the sessions
when the MSC-guided tool was made available to them, 84\% of the total tracing
time was spent using the tool on average across all users.
The ability to quickly switch between the MSC-guided and manual tools was reported
to be valuable when resolving topological issues.

Four of five subjects reported the MSC-guided tool was more comfortable to use
and allowed them to focus more on the data, as the tool requires less precise physical interaction
than manual tracing. Subject 4 did not report a significant perceived difference in terms of interaction effort. 
%
All users found the MSC tool's live preview trace valuable to review
the trace that would be selected before selecting it,
allowing for proofediting the trace on the fly during tracing.
The interactive aspect of the preview was reported to be particularly valuable when visualizing possible paths at intersecting neurons.

Four of five subjects reported that the flashlight feature was valuable for navigating 
the data, subject 2 did not find feature useful. Subjects 1 and 5, both experts in their field,
commented that the local preview provided by the flashlight was especially helpful when tracing
in poorly imaged regions and determining the endpoints of branches.
The zoom functionality was also found to be useful when navigating through the data,
with users zoomed out for the majority of time spent navigating.





\paragraph{Manual vs.\ MSC-guided Tracing.}
In our initial evaluation, the majority of users found the MSC-tool more challenging to learn.
A frequent comment we received was that it took a few traces to become
comfortable with the tool and to learn the cases where \revision{the tool would follow
the desired path and where it would not.} As a result of these comments,
we extended our training process to the one described in \Cref{sec:expert-eval} 
to help users familiarize themselves with the new technique.
With this new training in place,
we no longer received comments from users about the tool being difficult to use.
Although some users still reported becoming more comfortable with the tool over
time, the difference in tracing performance was not as significant.

Users reported the MSC-guided tool to be especially useful when tracing long axons
through large portions of the volume. When using the MSC tool they would
let it follow the neuron for them, and focus on navigating to the end point
of the axon to finish the trace.
The ability to zoom out and cover larger portions of the volume when tracing with
the MSC-tool accelerated this process significantly over manual tracing.
When tracing manually this task is
more difficult, as users must frequently swap between tracing and navigating
to create an accurate trace.
During the manual portion of the second session, many subjects lamented
being unable to use the MSC-guided method in these sections.

There are a number of regions in the data where a neuron
may appear to end or fade due to issues with the tissue labeling or
imaging process. All subjects reported that the MSC-guided tool
was particularly helpful in resolving these portions of the data.
One of the senior neuroanatomists, subject 1, reported
that using the MSC-guided tool helped him analyze these
cases more carefully. On one trace this led to him ultimately determining that
a neuron did not end where he initially thought it did.
Subjects \snine and \sten made similar comments, noting that
the MSC-guided tool and the flashlight preview helped them make decisions at 
potential branch and termination points.

Finally, during a visual inspection of the traces we found that in many
cases the MSC-guided trace followed the neuron more closely
than the DIADEM and VR reference traces (e.g.,~\Cref{fig:visual-msc-vs-diadem-vr,fig:all_trace_comparison,fig:man_deviation}).
The DIADEM trace clearly shows the limitations
of tracing manually on the desktop, where users click and place points
to construct straight line segments between them. 
The manual traces performed in VR, while producing a more refined line, show the difficulties of maintaining
the precise hand motion required to follow the neuron center accurately.
The MSC-guided tool alleviates the need for such precision by automatically following the ridge line
of the neuron, requiring the user to only provide a coarse set of inputs.

\revision{
\paragraph{Comparison to Desktop Software.}
Expert 5 has extensive experience using Imaris}~\cite{imaris_web} \revision{in his usual workflow. Imaris is a proprietary
software package for microscopy image analysis that provides a semi-automatic tracing feature with similar functionality.
When asked to compare the two methods, he reported that our MSC-guided tracing algorithm performed similarly to Imaris.
Moreover, he reported that the flashlight feature, which Imiris does not have, was valuable in resolving low resolution
and poorly imaged regions of the data. He also noted that the 3D navigation and interaction capabilities used in VR made the
tracing experience easier and more intuitive, compared to the large number of view point adjustments
that must be made when tracing on a desktop.}

\section{Summary and Future Work}
We have presented a novel semi-automatic neuron tracing method based on 
the topological features extracted from the Morse-Smale complex. We implemented our
MSC-guided tracing
tool within an existing VR environment, and demonstrated that it
improves neuron tracing performance over manual tracing in VR and semi-automatic methods
on a desktop. When using our MSC-guided tracing tool, experts were able to produce
acceptable quality traces with less fatigue and in far less time.
By leveraging the fast online computation time of our method, we are able
to show a live preview of the trace to the user, removing the need for
extensive post-process proofediting.
The neuroanatomists' qualitative feedback indicates
that, although more work remains to be done, our MSC-guided tool
is a promising approach to accelerate neuron tracing, especially
when tracing long range connections or in poorly imaged regions
and ambiguous regions.


Through iterating on the system design and training process with
experts, we have improved both the usability and interpretability
of our MSC-guided tool. The flashlight and preview features work together
to guide the neuroanatomist through the data and aid their decision making process.
At the same time, the zoom feature enables them to trace through the data faster.
Finally, our improved training process helps users get up to speed with the system faster to use the tool effectively when tracing.

A major challenge faced in our work were the structural and geometric differences in traces considered ``reference''.
The MSC-guided tracing method significantly reduces the geometric variation between traces, enabling the design of better tools for evaluating where expert traces disagree.
The high quality and consistent traces produced could be used to build a training set for labeling portions of an MSC-graph or image volume in a machine learning approach to advance automated neuron computation.  
Although the results of our pilot study are promising, we have also
found areas for improvement. Users expressed interest in seeing
additional information beyond the flashlight, which may make the tool
more informative for new and experienced users.
In addition, the trace previews and flashlight could be augmented by using a certainty measure to colormap the preview.
\section*{Acknowledgements}
The authors would like to thank Pavol Klacansky for assistance in the initial integration of the tool.
This work was funded in part by NSF OAC awards 1842042, 1941085,
NSF CMMI awards 1629660, LLNL LDRD project SI-20-001.
This material is based in part upon work supported by the Department of Energy (DoE),
National Nuclear Security Administration (NNSA), under award DE-NA0002375.
This research was supported in part by the Exascale Computing Project (17-SC-20-SC),
a collaborative effort of the DoE Office of Science and the NNSA.
This work was performed in part under the auspices of the DoE by Lawrence Livermore National
Laboratory under contract DE-AC52-07NA27344.
This work is supported by in part by grants from the NIH (R01 EY026812, R01 EY019743, BRAIN U01 NS099702), the NSF (IOS 1755431, EAGER 1649923), and The University of Utah Neuroscience Initiative, to A.A., a grant from Research to Prevent Blindness, Inc. and a core grant from the NIH (EY014800) to the Department of Ophthalmology, University of Utah.

\bibliographystyle{abbrv}

\bibliography{msc-vrnt,will-auto-export}

\end{document}